\documentclass[aps,prl,preprint,groupedaddress]{revtex4}
\usepackage{amsmath,amssymb}
\usepackage{amsbsy}
\usepackage[utf8]{inputenc}
\usepackage[english]{babel}
\usepackage{graphicx}
\usepackage{subfigure}
\newtheorem{theorem}{Theorem}
\newcommand*\DAlambert{\mathop{}\!\mathbin\Box}
\usepackage[a4paper,bindingoffset=0.2in,%
            left=.2 in,right=.2 in,top=1in,bottom=1in,%
            footskip=.25in]{geometry}

\def\bea{\begin{eqnarray}}
\def\eea{\end{eqnarray}}
\def\be{\begin{equation}}
\def\ee{\end{equation}}
\def\nn{\nonumber}

\def\p{\partial}
\begin{document}

\title{Scalar and Fermion field interactions with a gravitational wave}
\author{Siddhartha Morales $^{1, }$ $^2$}
\author{Arundhati Dasgupta$^1$
}

\affiliation{$^1$ Physics and Astronomy, University of Lethbridge, 4401 University Drive, Lethbridge, Canada T1K 3M4.}
\email{arundhati.dasgupta@uleth.ca}

\affiliation{$^2$ Facultad de Ciencias, UNAM
Ciudad Universitaria, D.F.
Mexico}
\email{sidd.morales@ciencias.unam.mx}
\author{}
\affiliation{}

\begin{abstract}

Given that gravitational waves are the future probes of the universe, it is important to test various physical effects of these on matter. This article explores the first-order perturbation, caused by a planar gravitational wave, on massless scalar and chiral fermions. We find a new propagating mode for the perturbed fields with a new dispersion relation by solving inhomogeneous differential equations. We also discuss the massive scalar field, and find interesting effects due to the gravitational wave.
 Our results have physical implications for early universe cosmology and also for ground-based observations, where the new mode might be able to help in the gravitational waves' detection on earth. \end{abstract}


\maketitle

\section{Introduction}
With the first detection of gravitational waves at the LIGO detector in 2015, and later announcement in 2016, a new era of observational physics has started \cite{ligo,ligo2}. The gravitational wave will soon become a major probe of the universe and it's dynamics. In this article, we discuss the question: how does a gravitational wave affect the behaviour of a scalar wave? This scalar wave could be a matter wave in a condensed matter system, or could represent the inflaton of the early universe \cite{cosmo}. Gravitational waves are important in the dynamics of the early universe, though they are expected to be non-linear and at a much higher amplitude than that detected on earth recently \cite{ligo}. In our calculations, we test whether a linear gravitational wave can interact with a scalar field to give non-trivial dynamics.  The focus is in the way the matter waves (scalars and fermions) react to the gravitational wave. Whereas we have done a search for papers which report a similar scalar field gravitational wave interaction \cite{singleton}, our calculations are unique. In particular, we seem to have obtained a `non-trivial' response of the massless scalar wave with the gravitational wave, which we discuss and try to obtain new physical implications. In a similar calculation in \cite{singleton} additional restrictions are imposed on the scalar field to obtain a planar wave, here we do not impose such restrictions. In our paper the scalar field responds to the passing gravitational wave, and is perturbed to an oscillatory reaction  evident in the plot of the energy of the perturbation.  If we take a gravitational wave in the $z$ direction,  then in the perturbation equations,  the inhomogeneous terms are proportional to $k_{0x}^2 - k_{0y}^2$  and $k_{0x} k_{0y}$ (where $k_{0x}, k_{0y}$ are the $x-y$ components of the scalar field's direction of propagation). Thus, at least one of the transverse directions of the scalar wave vector has to be non-zero for the gravitational wave in the $+z$ direction to have any non-trivial effect on the scalar wave. The spherical propagator used to obtain  the scalar wave perturbation is also crucial for the results, if the scalar waves are obtained using the planar wave propagator as we show in the massive scalar example, the plane waves are perturbed  as ordinary waves. Our solutions, are obtained using Kirchhoff's Theorem and Duhamel's principle, and the separation of variables method is not used to solve the Laplacian. The new result is that the perturbations have a new mode, obtained using the spherical propagator for the solution to the inhomogeneous wave equation. The new mode can be interpreted as massless waves of frequency $\omega$ with the dispersion $\omega^2= (\textbf {k}+ \textbf {k}_0)^2$, where $\textbf k$ is the wave vector of the gravitational wave, and $\textbf k_0$ is the wave vector of the scalar/fermion wave.  We check for resonant modes, in the solutions for the scalar field as has been found in `driven systems' and find none.

Similarly we investigate the effect of the gravitational wave on a chiral fermion (neutrino), and obtain the first order effect on the particle.  Similar to the scalar particle,  our solutions have different behaviour not seen previously in \cite{fermion}. The neutrino has a non-trivial interaction with the gravitational wave and is perturbed to an oscillatory flux flow. This result will have physical implications, in particular, we can expect corrections to the effective number of neutrino species for the cosmic neutrino background. The effective number of neutrino species is $3.15 \pm 0.23$ \cite{neutspecies}, close to the physically expected 3.0 $\pm$ corrections, the fluctuations have been attributed to non-thermal interactions in the early universe \cite{neut1,neut2}. In the section \textit{Field Perturbations for massless spin $1/2$ particles}, we estimate a correction due to the gravitational wave interaction, and the perturbation effect's contribution to the background cosmic neutrino density. The perturbation of the density is very small to contribute to the effective degrees of freedom, however the neutrino interaction with the cross polarization has a resonant mode. On page 18, and using equation (\ref{eqn:modes}) we comment on how the resonant run away mode can contribute to the background density of neutrinos.

 These perturbations might also affect the neutrino flux for the multi-messenger astronomy observations from neutron star collisions. As of now neutrinos are yet to be detected from the gravitational wave events. More importantly, the fluctuations in the neutrino flux produced due to the gravitational wave interaction can be used to detect gravitational waves on earth. Particularly for fermions perturbed by the cross polarization of the gravitational waves, there is a resonant mode which grows with time (\ref{eqn:modes}). The physical implications of this have to be investigated in a real experimental situation in consultation with existing neutrino experiments \cite{adg}. This is work in progress and as of yet neutrinos have not been used to detect gravitational waves on earth. 
 
  \subsection{Comparison with previous studies of matter-gravitational wave interaction}
 We did a deep search for previous work on matter and gravitational waves interactions, and we analyze how our results compare to those. Our aims are very similar to the motivations of \cite{singleton, singleton2} for scalars and \cite{fermion} for the neutrinos. However, our results are different, as in \cite{singleton} the perturbed scalar waves have $\left(\partial^2_x-\partial^2_y \right)\phi=0$ to produce planar scattered waves.  We do not impose any such additional restrictions, on the perturbation, and thus our results are different. In \cite{fermion,pprod}, the scattered equation for the neutrinos are solved using separation of variables, and thus the particular mode found in this paper is not manifest. In other papers brought to our attention by reviewers, \cite{hog,bamba,cet,vald,vald2} the focus is mainly on the interaction of gravitational waves with electromagnetic waves.; and our calculations have different results. Our calculations discuss inhomogeneous scalar wave equations and Weyl equations and we solve to obtain a particular solution which we call a `new mode' for the fields. This analysis has not been done previously for electromagnetic waves where Maxwell's equations are solved in gravitational wave background. In particular in \cite{bamba} the scalar fields were studied, but these are quantized and are sources for the gravitational wave. In \cite{cet}, the calculations are computed of semiclassical interactions of matter with gravitational waves using field theoretic techniques, and the results are not the same as obtained in our paper, in particular in the scattered quantum wave, we do not find any mode with the same dispersion as in this paper. In the papers \cite{vald,vald2}, perturbations of flat space time are studied using the vierbein and the spin-connection, but they do not have a mode analysis, as in Equations (\ref{eqn:modesf1},\ref{eqn:modesf}) does, in \cite{vald2} as is presented in this paper. In \cite{vald,vald2}, the perturbations include torsion.
 Interactions of gravitational waves with matter has also been studied in \cite{grish} and \cite{chris}. The results obtained in Equations (\ref{eqn:masslesssc},\ref{eqn:modesf1},\ref{eqn:modesf},\ref{eqn:modes}) are new and different as discussed in these papers. The solutions found in the above equations have a new dispersion relation $\omega^2= (\textbf {k}+ \textbf {k}_0)^2$, where $\textbf k$ is the wave vector of the gravitational wave, and $\textbf k_0$ is the wave vector of the scalar/fermion wave, these have not been obtained in \cite{grish, hog,bamba,cet,vald,vald2,singleton,singleton2}. In particular in \cite{grish} , particles interact with gravitational waves, and there are similar `forced' oscillation equations which are solved. However, in contrast, those perturbations solutions of particle motion show resonance, whereas our scalar modes are non-resonant. Our paper also discusses the fermion scattering from gravitational waves which are not found in \cite{grish}. 
 
 The paper is organized as follows: in the next section, we describe the Klein-Gordon equation and its solution in the gravitational wave background. In the third section we describe the chiral fermion and its solution in the gravitational wave background. In the fourth section, we interpret the solution, and provide some practical uses of the solution, and we conclude.
\section{\label{sec: Klein Gordon} The Klein Gordon equation in the background of a gravitational wave \\} 

The Klein-Gordon equation in the background of a curved metric $g_{\mu \nu}$ is
\begin{equation}
\partial_\mu \left( \sqrt{\text{-}g} \  
 \partial^\mu \phi \right) =  \partial_{\nu} \left( \sqrt{\text{-}g} \ g^{\mu\nu} \partial_{\mu} \phi \right)= 0. 
\end{equation}

In this article we solve this equation in the background, where $g_{\mu \nu}$ is due to a gravitational wave. We take the example of a perturbation due to a gravitational wave of a flat geometry, which could be the situation on Earth,  when the gravitational wave arrives from a distant source.  
\subsection{\label{sec: Gravitational Wave} Gravitational Waves \\} 

The simplest kind of  gravitational waves emerges in the frame of the linearized gravity theory, obtained by approximating the metric tensor up to first order.  In the linear theory approximation, one can say that the metric is flat space-time with fluctuations $h_{\mu \nu}$ as 
\begin{equation}
g_{\mu\nu} = \eta_{\mu\nu} + h_{\mu\nu},
\end{equation}
where $\lVert h_{\mu\nu} \rVert << 1$. Using this, it can be proved that the inverse of the metric can be written as 
\begin{equation}
g^{\mu\nu} = n^{\mu\nu} - h^{\mu\nu}
\end{equation}
 (if we want to keep everything up to first order). \\

Introducing the ``traceless" metric 
\begin{equation}
\bar{h}_{\mu\nu} = h_{\mu\nu} - \frac{1}{2}\eta_{\mu\nu}h^{\alpha}_\alpha,
\end{equation}

Einstein's tensor becomes

\begin{equation}
G^{\beta \mu}=-\frac{1}{2} \left( -\partial_{t}^2+\bigtriangledown^{2} \right) \bar h^{\beta\mu}= \DAlambert \bar h^{\beta\mu}.
\end{equation}

Using gauge freedom we can choose Lorentz gauge $\bar h^{\lambda\sigma},_{\sigma}=0$, together with the so called Transverse-Traceless gauge (TT-gauge), which imposes on $\bar{h}_{\mu\nu}$ the following conditions: $\bar{h}_{\mu}^\mu = 0$, then $\bar{h}_{\mu\nu} = h_{\mu\nu}$ and $\partial^{\mu} h_{\mu\nu} = 0$. The metric is divergence-less $ h_{\mu\nu} u^\mu = 0$, for some 4-velocity $u$.
Imposing both gauges, Einstein's equation can be written as
\begin{equation}
\DAlambert \bar{h}_{\mu\nu} = -16 \pi T_{\mu\nu}
\end{equation}

(in units of $G=c=1$ or Newton's constant and speed of light set to 1). 
In the absence of matter, i.e. $T_{\mu\nu} = 0$, the last equation is just the three-dimensional wave equation, which has as solutions plane waves. Note, the gravitational waves are initially taken as in vacuum, and their effect is studied on the scalar fields. We are not studying back reaction of the matter fields on the gravitational waves, in that case $T_{\mu \nu}\neq 0$ and this is not the focus of the paper. 

\begin{equation}
\bar{h}_{\mu\nu} = A_{\mu\nu} \ \mathrm{e}^{i k_{\alpha}x^{\alpha}},
\end{equation}

where $A_{\mu \nu}$ is the tensor polarization amplitude. Also, since $h_{\mu\nu}$ is symmetric it has only $10- 4 -4 = 2$ independent components.
We can set the coordinate axes so that the gravitational wave is propagating in the $z$ direction. With these conditions and the TT-gauge, the linearized metric tensor can be expressed as
\begin{align}
g_{\mu\nu} = \begin{pmatrix}
    -1 & 0 & 0 & 0 \\
    0 & 1+h_+ & h_\times & 0 \\
    0 & h_\times & 1-h_+ & 0 \\    
    0 & 0 & 0 & 1
  \end{pmatrix}    
 \,\,\,\,\,\, \ ; \ \ \ \ \ \ \ \ 
 g^{\mu\nu} = \begin{pmatrix}
    -1 & 0 & 0 & 0 \\
    0 & 1-h_+ & -h_\times & 0 \\
    0 & -h_\times & 1+h_+ & 0 \\    
    0 & 0 & 0 & 1
  \end{pmatrix}  
\end{align}

(where $h_+ = A_+ \cos(\omega(z-t))$ and $h_{\times}=A_\times \cos(\omega(z-t)+ \delta)$, $\omega$ the frequency of the wave and $A_{+}, A_{\times}$ being the amplitudes of the two polarizations) 

Let $g$ be the determinant of $g_{\mu\nu}$, we can see that due to the block diagonal form of the metric, the determinant is
\begin{equation}
g = -(1 - h_{+}^2 - h_{\times}^2) \approx -1 
\end{equation}
which is approximately $1$ up to first order.

For numerical estimates; e.g. binary white dwarf systems in our galaxy, a good approximation of the wave strain is \cite{flan}

\begin{equation}
h \approx 10^{-21} \left( \frac{M}{2M_\odot} \right)^{5/3} \left( \frac{1 \, \mbox{hour}}{P} \right)^{2/3} \left( \frac{1 \, \mbox{Kpsc}}{r} \right).
\end{equation}

For binary neutron stars spiralling together in our galaxy, a good approximation of the wave strain is \cite{flan}

\begin{equation}
h \approx 10^{-22} \left( \frac{M}{2.8 M_\odot} \right)^{5/3} \left( \frac{0.01 \, \mbox{sec}}{P} \right)^{2/3} \left( \frac{100 \, \mbox{Mpsc}}{r} \right)
\end{equation}

where $P=2\pi/\Omega$, $\Omega$ is the orbital frequency and $r$ is the distance from the source. M is the total mass and $M_{\odot}$ is the mass of the sun. Similarly, quasi-normal ring down of stellar mass black hole mergers are expected to give waves of amplitudes \cite{ligo2}
\begin{equation}
h \approx 10^{-21} \left(\frac{\nu}{0.25}\right) \left(\frac{M}{20 M_{\odot}}\right)\left(\frac{200 {\rm Mpsc}}{r}\right)
\end{equation} 
($\nu$ is the symmetric mass ratio \cite{ligo2}). 

\section{\label{sec: Wave equation in 3d} The interaction of the scalar wave and the gravitational wave \\} 

Let's say that we have a field satisfying Klein-Gordon's equation for a massless particle in a metric produced by a gravitational wave.
The equation for the scalar field in a curved space is given by $\partial_\mu \left( \sqrt{\text{-}g} \partial^\mu \phi \right) =  \partial_{\nu} \left( \sqrt{\text{-}g} g^{\mu\nu} \partial_{\mu} \phi \right)= 0 $, and remembering that the determinant for the metric of a gravitational wave is equal to $-1$, the equation is just $$\partial_{\nu} \left(  g^{\mu\nu} \partial_{\mu} \phi \right)= 0.$$ 
Expanding all components in the equation and noting that $g_{\mu\nu}$ has only six non-zero components and that $h_{\mu\nu}$ is a function only of $z$ and $t$, the equation is explicitly 
\begin{equation}
0 = -\partial_{tt} \phi +\partial_{xx} \phi +\partial_{yy} \phi +\partial_{zz} \phi - \left[ h_+ ( \partial_{xx} \phi - \partial_{yy} \phi ) + 2h_{\times} \partial_y \partial_x \phi \right]
\label{eqn:scalar}
\end{equation}
or in a more compact form
\begin{equation}
\DAlambert \phi - \left[ h_+ ( \partial_{xx}  - \partial_{yy} ) + 2h_{\times} \partial_y \partial_x  \right] \phi=0.
\end{equation}

Any two variable function can be expanded around two variables, in this case $A_+$ and $A_\times$. Due to the smallness of the strain of a gravitational wave $(10^{-21}-10^{-22})$, we can make a Taylor expansion up to first order in $A_+$ and $A_\times$ as continuous variables, thus this result would be
\begin{equation}
\phi \approx \phi_0 + A_+ \phi_+ + A_\times \phi_\times,
\end{equation} 

where $\phi_+ = \frac{\partial }{\partial A_+}\phi$ and $\phi_\times = \frac{\partial }{\partial A_\times}\phi$.

Substituting this expression into the Klein-Gordon with the metric of a gravitational wave, and keeping terms up to first order gives
\begin{eqnarray}
 \DAlambert \phi_0 - \left[ A_+ \cos(\omega(z-t)) (\partial_{xx} \phi_0 - \partial_{yy} \phi_0) + 2A_{\times}\cos(\omega(z-t) + \delta) \partial_x \partial_y \phi_0 \right] && \nonumber \\+ A_+ \DAlambert \phi_+ + A_\times\DAlambert \phi_\times&=&0.
\end{eqnarray}

Since $A_+$ and $A_\times$ are arbitrary constants we get the following set of equations
\begin{align}
\DAlambert \phi_0 &= 0, \ \ \ \mbox{(unperturbed wave equation) } \\
\DAlambert \phi_+ &= \cos(\omega(z-t)) (\partial_{xx} - \partial_{yy})\phi_0 \\
 \DAlambert \phi_{\times} &= 2\cos(\omega(z-t)+\delta) (\partial_{x} \partial_{y})\phi_0
\end{align} 

For sake of simplicity let's suppose that 
\begin{equation}
\phi_0 (\textbf{x}, t)= \Re \left( A_0\mathrm{e}^{i(\textbf{k}_0 \cdot \textbf{x} - \omega_0 t)} \right)
\end{equation}

(Notation: we use boldface to represent three vectors)

The perturbations are next solved for using techniques of inhomogeneous PDE's.

\subsection{The wave equation in 3+1 d}

The wave equation in $\mathbb{R}^3$ is just the D' Alambertian of a function equated to zero, $\DAlambert \phi = 0$. Further, let's suppose that we have Cauchy (initial) conditions, 
\begin{align}
\DAlambert \phi &= 0 \\
\phi(\textbf{x},0) &= F(\textbf{x}) \\
\frac{\partial \phi}{\partial t} (\textbf{x},0) &= G(\textbf{x})
\end{align}
where $\textbf{x} \ \in \mathbb{R}^3$, $t>0$, $ F  \in C^3(\mathbb{R}^3)$ and $G \ \in \ C^2(\mathbb{R}^3)$. 

The solution for the wave equation with Cauchy conditions is given by the theorem \cite{kirch}.

\begin{theorem}
Suppose that $ F  \in C^3(\mathbb{R}^3)$ and $G \ \in \ C^2(\mathbb{R}^3)$. Then the solution of the initial value problem of the wave equation in three dimensions is given by
\begin{equation}
\phi(\textbf{x},t) = \frac{1}{4\pi t} \int_{S(\textbf{x},t)} G(\textbf{x}') d \sigma_t + \frac{\partial}{\partial t} \left[ \frac{1}{4 \pi t} \int_{S(\textbf{x},t)} F(\textbf{x}') d \sigma_t \right]
\end{equation}
and the solution is in $C^2$ for $\textbf{x} \in \mathbb{R}^3$ and $t \leq 0$.
\end{theorem}

The above formula is called Kirchhoff's formula.

The solution for the inhomogeneous wave equation is given by \textit{Duhamel's principle}, which says that if $\phi(\textbf{x},t)$ is the solution of the following initial value inhomogeneous wave equation, \cite{pde}
\begin{align}
\DAlambert \phi = h(\textbf{x},t); \ \ \ \textbf{x} \in \ \mathbb{R}^3, \ \ \ t > 0, \\
\phi(\textbf{x},0) = 0, \ \ \ \ \frac{\partial \phi}{\partial t}(\textbf{x},0) = 0; \ \ \textbf{x} \in \ \mathbb{R}^3.
\end{align}
Let $v(x,t;\tau)$ be the solution of the associated ``pulse problem"
\begin{align}
\DAlambert v = 0; \ \ \ \textbf{x} \in \mathbb{R}^3, \ \ \ t > \tau, \\
v(x,\tau;\tau) = 0, \ \ \ \frac{\partial}{\partial t} v(x,\tau:\tau) = - h(\textbf{x},\tau)
\end{align}
Then the solution for the inhomogeneous wave equation is given by
\begin{equation}
\phi(\textbf{x},t) = \int_{0}^t v(x,t:\tau) \ d \tau
\end{equation}

Combining Duhamel's principle and Kirchhoff's formula, the solution for the inhomogeneous wave equation in three dimensions can be written in compact form as
\begin{equation}
\phi(\textbf{x},t) = -\frac{1}{4\pi} \int_{\bar{B}(\textbf{x},t)} \frac{h(\textbf{x}',t-r)}{r}d^3 x'
\end{equation}
where $r = \lVert \textbf{x} - \textbf{x}' \rVert = \sqrt{(x-x')^2+(y-y')^2 + (z-z')^2}$ and $\bar{B}(\textbf{x},t)$ is the ball in $\mathbb{R}^3$ with center at $\textbf{x}$ and radius $t$.

The solution of $\phi_+$ and $\phi_{\times}$ would be
\begin{align}
\phi_+ (\textbf{x},t) &= \frac{1}{4\pi} (k_{0x}^2 - k_{0y}^2) \int_{\bar{B}(\textbf{x},t)} d^3 \textbf{x}' \frac{\phi_0 (\textbf{x}',t')}{r} \cos(\omega(z'-t')) \\
\phi_\times (\textbf{x},t) &= \frac{1}{4\pi} (2k_{0x}k_{0y}) \int_{\bar{B}(\textbf{x},t)} d^3 \textbf{x}' \frac{\phi_0 (\textbf{x}',t')}{r} \cos(\omega(z'-t')+\delta)
\end{align}
where $\textbf{k}_0 = (k_{0x},k_{0y},k_{0z}) $, $r = \lVert \textbf{x}' - \textbf{x} \rVert$ and $t' = t - r$.

Now, in order to find the perturbation explicitly, let's solve the more general integral.
\begin{equation}
I(\delta_0,\delta) = \int_{\bar{B}(\textbf{x},t)} d^3 \textbf{x}' \ \mathrm{e}^{i(\textbf{k}_0 \cdot \textbf{x}' - \omega_0 (t') + \delta_0 )}  \frac{\cos(\omega(z'-t') + \delta)}{r}
\end{equation}
where $r = \lVert \textbf{x}' - \textbf{x} \rVert$ and $t'=t-r$. Since we are integrating over the ball centered at $\textbf{x}$ we can make a change of variable (reference system) and center the ball at the origin, the translation would be
\begin{align}
\textbf{x}' &= \textbf{x} + \textbf{r}  \label{eqn:define1}\\
d^3 \textbf{x}' &= r^2 \sin(\theta) d\phi d\theta dr. \nonumber
\end{align}

With the last change of variable and writing the cosine term in its exponential form, then the integral is converted into
\begin{align}
I &= \int_{\bar{B}(0,t)} \mathrm{e}^{i(\textbf{k}_0 \cdot (\textbf{x} + \textbf{r}) - \omega_0(t-r)+\delta_0)} \frac{\mathrm{e}^{i(\omega(z' - (t-r)) + \delta)}+\mathrm{e}^{-i(\omega(z' - (t-r)) + \delta)}}{2r} r^2\sin\theta d\phi d\theta dr 
\end{align}

This integral can be done using \cite{gradshteyn} and Mathematica (explicit steps are shown in appendix 1).  The result is:

\begin{align}
&\mathrm{e}^{i(\textbf{k}_0 \cdot (\textbf{x}) - \omega_0(t)+\delta_0)} \times \nonumber\\
&\times \frac{\pi \mathrm{e}^{i(\omega(z - t) + \delta)}}{\lVert \textbf{k} + \textbf{k}_0\rVert} \frac{ \left(-\lVert \textbf{k} + \textbf{k}_0\rVert+e^{i t \left(\omega +\omega _0\right)} \left(\lVert \textbf{k} + \textbf{k}_0\rVert \cos (\lVert \textbf{k} + \textbf{k}_0\rVert t)-i \left(\omega +\omega _0\right) \sin
   (\lVert \textbf{k} + \textbf{k}_0\rVert t)\right)\right)}{\omega \omega_0 (1-\cos\beta)} + (\omega\rightarrow -\omega)
\end{align}
 In the above equation $\beta$ is the angle between the wave vectors of the scalar wave and the gravitational wave in three dimensions, $\beta = \frac{\textbf{k}\cdot \textbf{k}_0}{\lVert \textbf{k} \rVert \lVert \textbf{k}_0\rVert}$. Also, the added term $(\omega \rightarrow -\omega)$  is the expression already shown in the above equation, with only the change that $\omega$ is replaced by $-\omega$.
 
 From the above one can see that apart from the usual super-position of two waves where the frequencies add or subtract, there is a term of the form $\cos(|| \textbf{k} + \textbf{k}_0|| \ t)$.
 To analyze the results, we write the waveform in the following way:
 \begin{equation}
 \frac{\pi}{ \omega\omega_0(1-\cos\beta)} e^{\left(i \left(\textbf{k}_0+\textbf{k}\right)\cdot\textbf{x}\right)} \left[-e^{-i(\omega+\omega_0) t} +   \left[ \cos(||\textbf{k}+\textbf{k}_0||  ~ t) - i \frac{(\omega+\omega_0)}{||\textbf{k}+\textbf{k}||}\sin(||\textbf{k}+\textbf{k}_0||~t)\right]\right] + (\omega\rightarrow -\omega)
 \label{eqn:masslesssc}
 \end{equation}
 
\begin{center}
\begin{figure}
\subfigure{\includegraphics[scale=0.3]{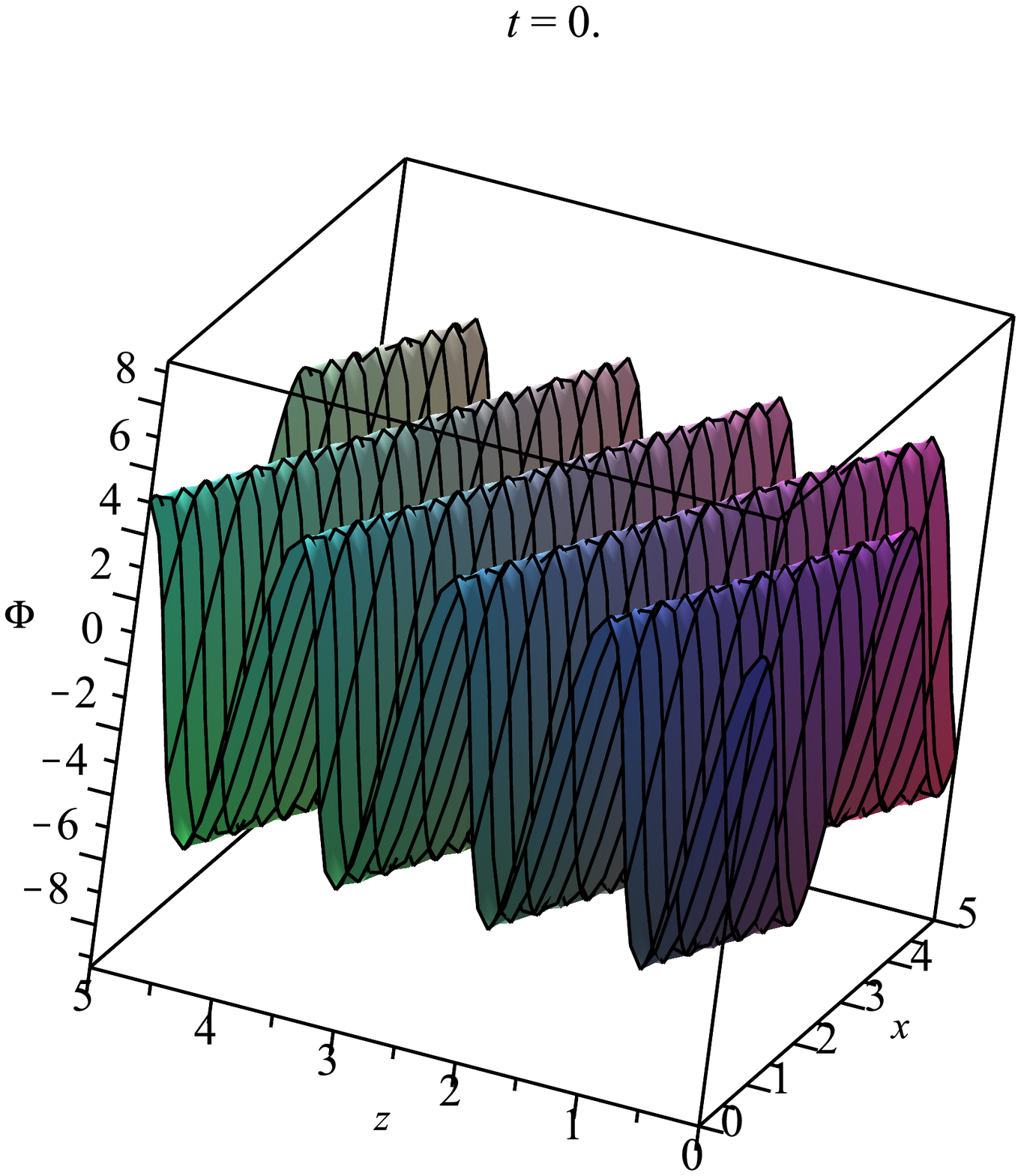}}
\subfigure{\includegraphics[scale=0.3]{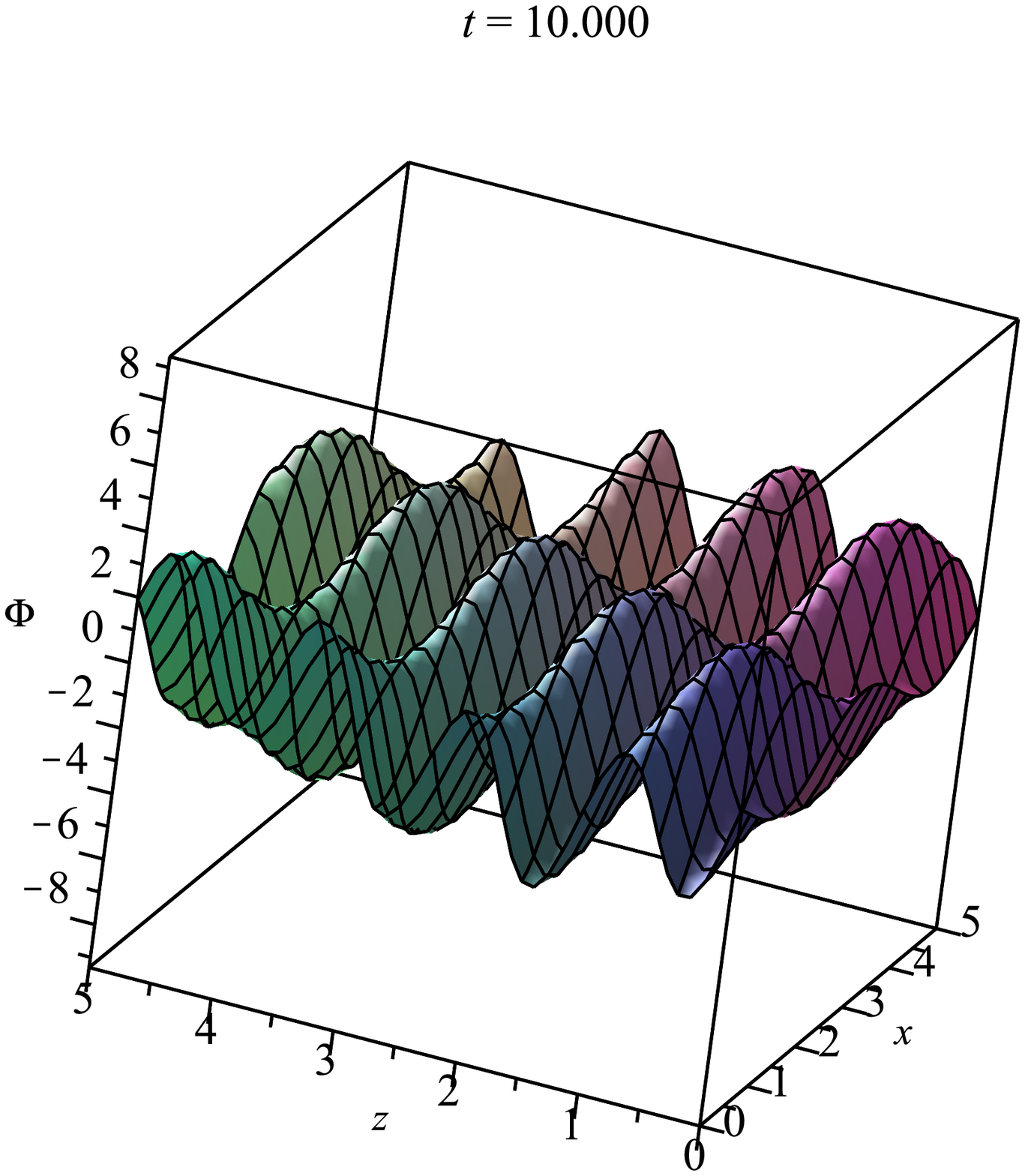}}
\subfigure{\includegraphics[scale=0.3]{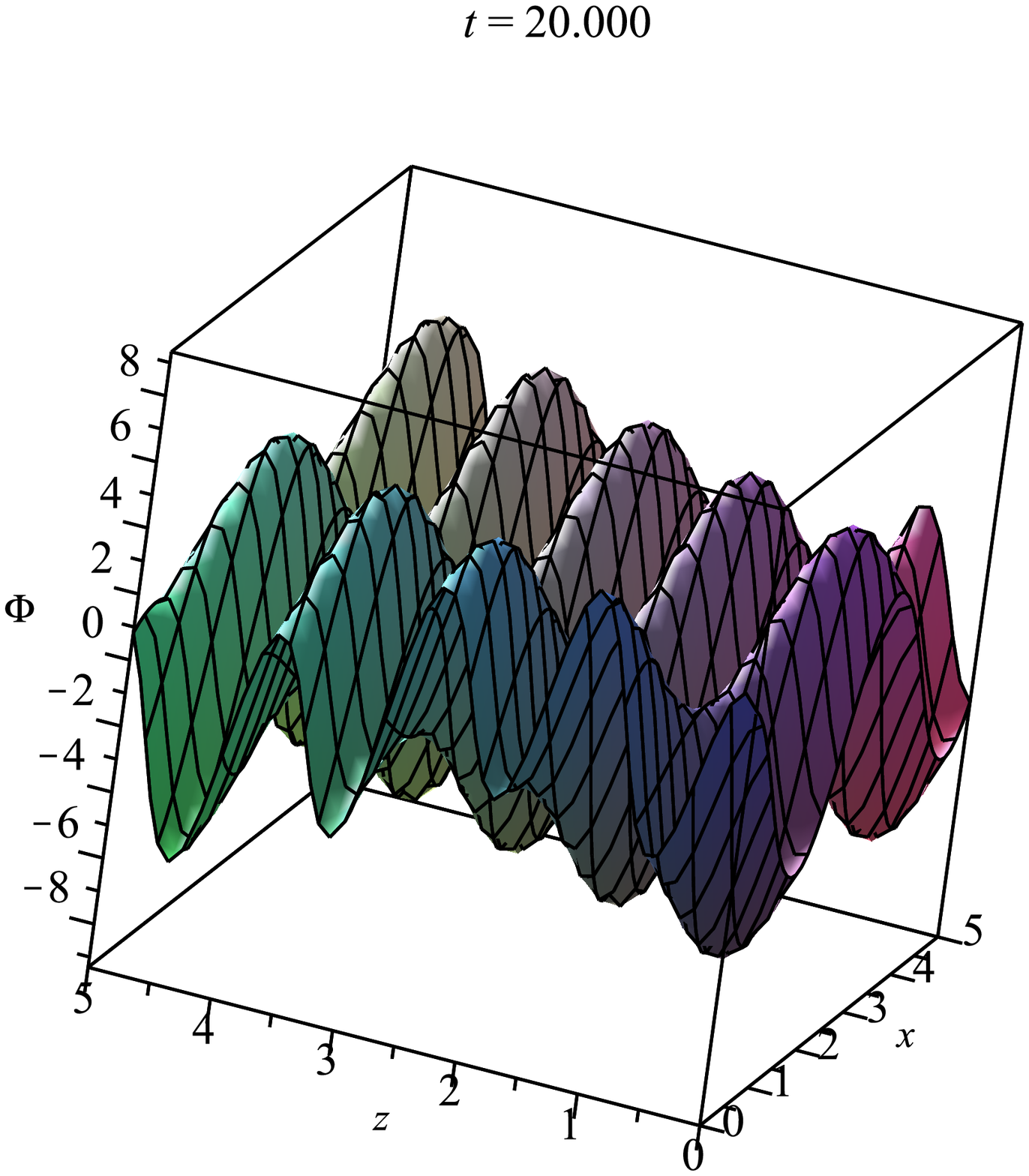}}
\subfigure{\includegraphics[scale=0.3]{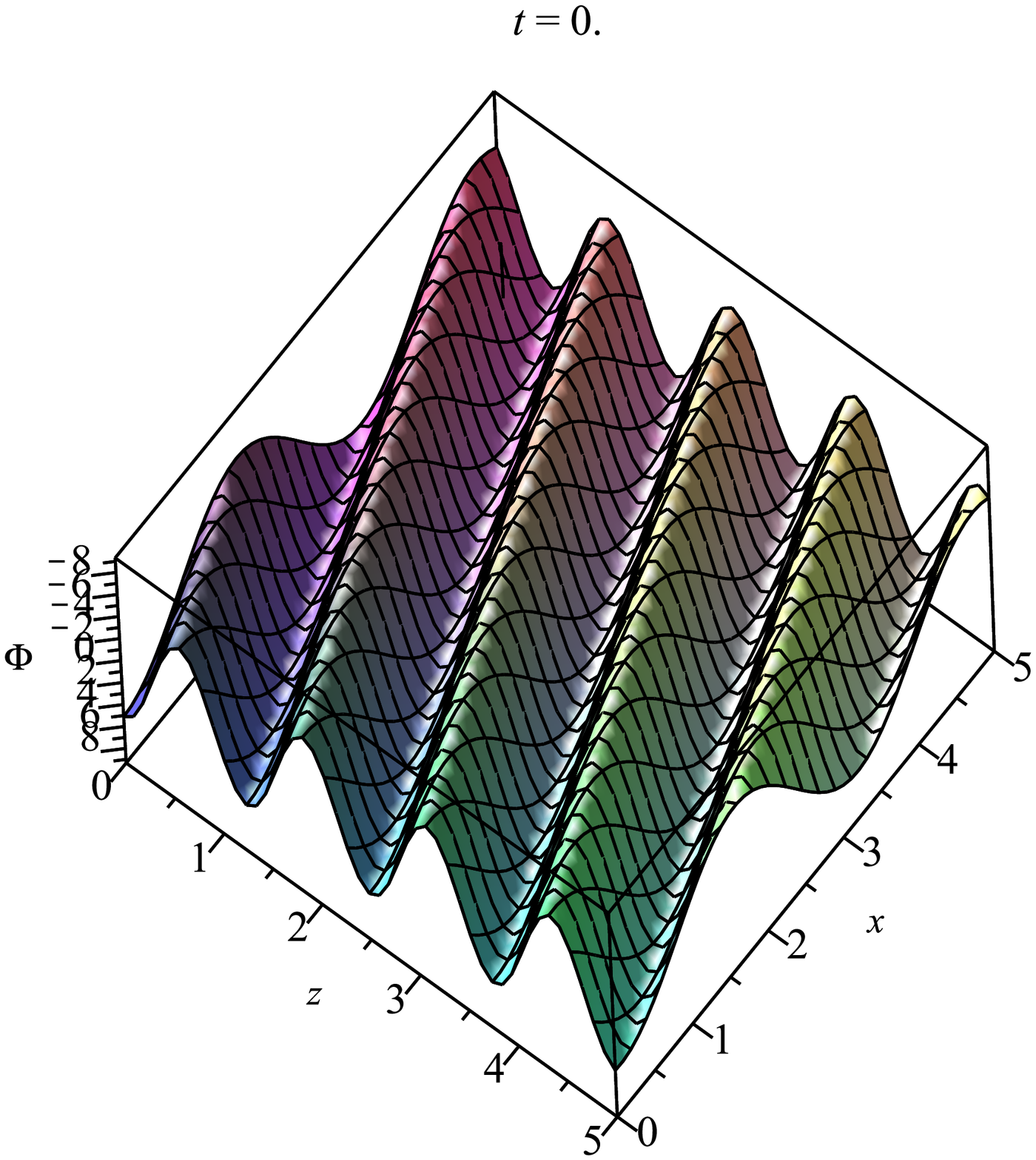}}
\subfigure{\includegraphics[scale=0.3]{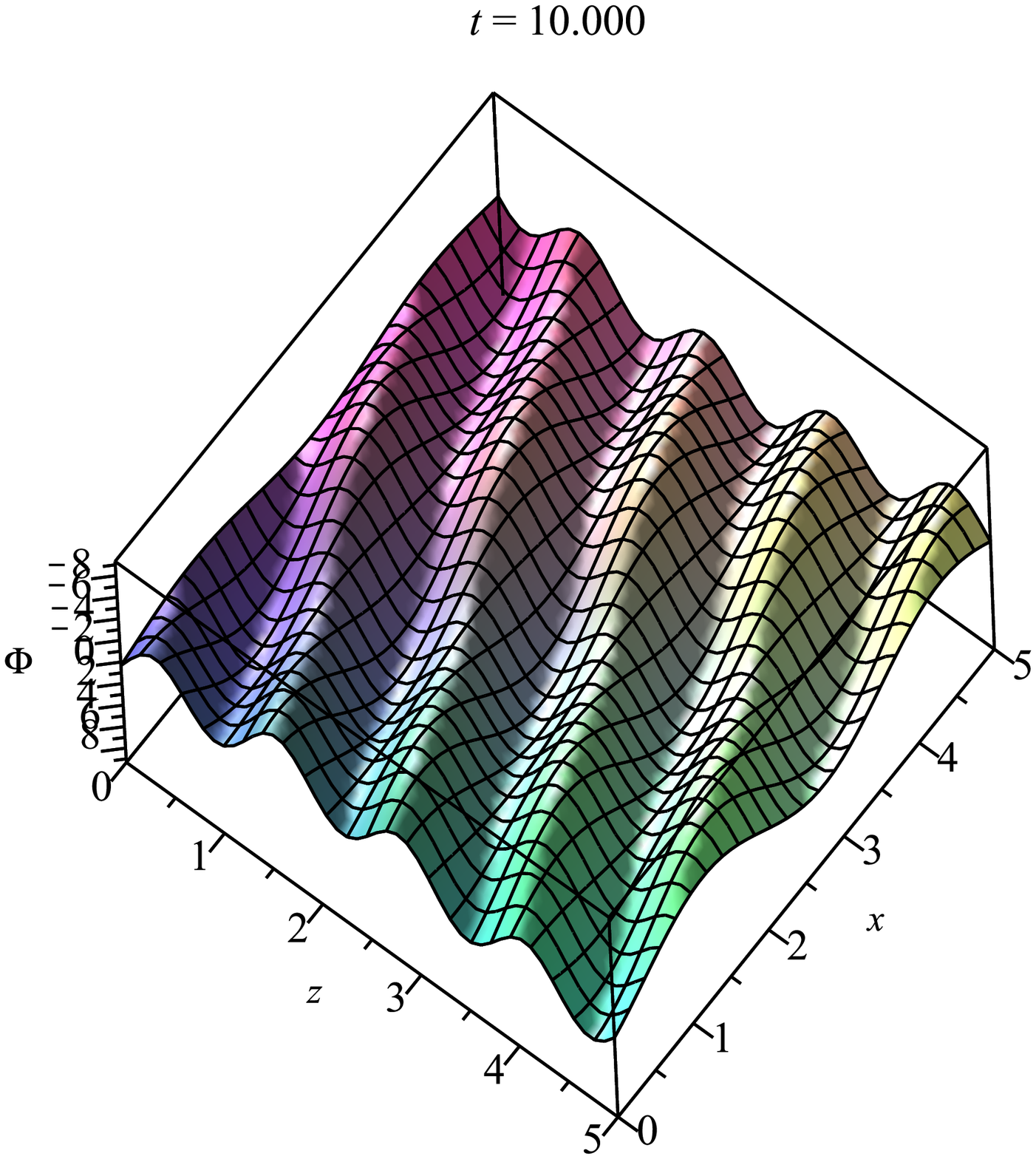}}
\subfigure{\includegraphics[scale=0.3]{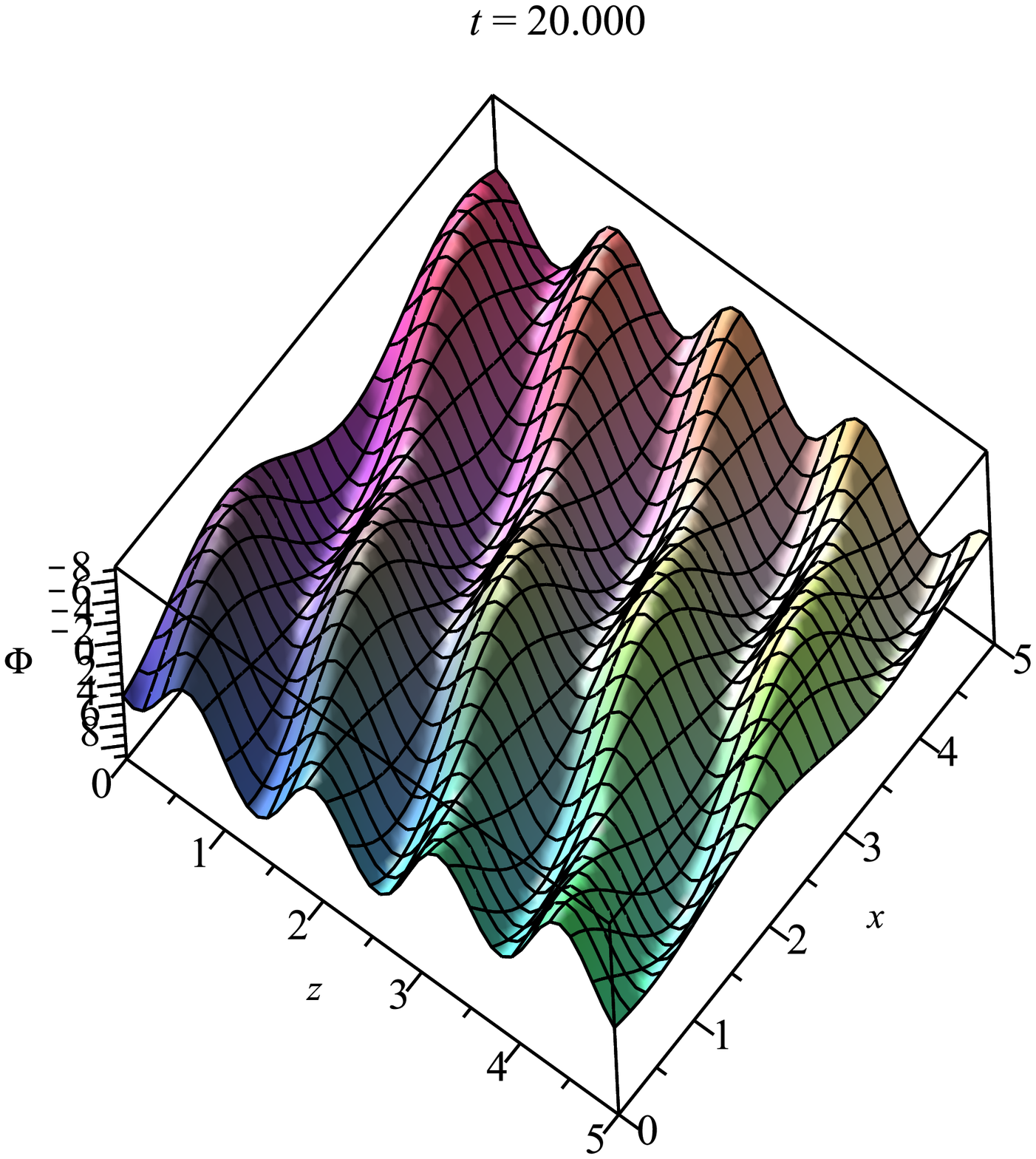}}
\caption{The perturbed scalar field (real part) at a sequence of time steps, which show oscillation in time as a response to the gravitational wave plotted in 3 d for two different perspectives and x and z ranges. For the purpose of plot, the $k_{0x}=k_{0z}=2, k_z=3$}
\label{fig:Scalar}
\end{figure}
\end{center}

A plot of the above using MAPLE shows the behaviour of the solution in time. The first part comprises of an ordinary part of frequencies $\omega\pm\omega_0$, and the second part comprises of a wave of  frequency  $||\textbf{k}+\textbf{k}_0|| $, as in Figure (\ref{fig:Scalar}).
 
 The solution does not show resonance, as if we set $\beta=0$ (the angle in momentum space between the gravitational wave and the scalar wave), which will make the amplitude singular, the solution itself is zero.
Despite this, the solution shows the promise that waves produced by spin 0 particles (described by scalar fields as in equation (\ref{eqn:scalar}))  can be used to detect the passing of a gravitational wave on earth due to the solution of equation (\ref{eqn:masslesssc}).  Note that the unperturbed wave does not show any amplitude oscillation, and is simply propagating with constant energy. The presence of the gravitational wave causes interesting oscillation patterns as in Figure(\ref{fig:Scalar}) albeit these perturbations will be very weak.
The physical reason for the oscillating amplitude is of course the `energetic interaction' of the scalar wave with the gravitational wave in response to the stretching and straining of the space-time through which it propagates. Note that scalar waves donot exist but this result could be true for Electromagnetic (EM) waves, as each component of the EM wave satisfies a scalar wave equation though the details of EM-gravitational wave are work in progress). In the early universe, the inflaton or scalar perturbations could be a source of `scalar waves'. 
To estimate the amplitude of the perturbation let's say that the gravitational wave has only $+$ polarization and that $\textbf{k}_0$ propagates only in the $x$ direction, thus
\begin{align}
\Delta \phi & \approx A_{+} \phi_{+} \sim h \phi_+ \\
& \sim  h \frac{A_0 k_{0}^2}{2\omega\omega_0} \sim hA_0 \left( \frac{\omega_0}{\omega} \right)
\end{align}

As a result, the order of the perturbation is proportional to $ h \left( \frac{\omega_0}{\omega} \right)$, where $h$ is the gravitational wave strain. We already know that this strain is of the order of $10^{-21} - 10^{-22}$, and for $ \omega_0 = 160.23 Ghz$, the peak CMB frequency as an example;  \footnote{Here we have taken the peak frequency of the CMB as an example; one component of the EM field satisfies similar inhomogeneous wave equation; (work to appear)} and $\omega \in (10^{-7} Hz -100 Hz)$, the amplitude of the perturbation would be in the range of $10^{-13} - 10^{-3}$.
Even if it is weak compared to the scalar wave, which we take as order 1, it is certainly detectable, the frequency of oscillation would be however of the order of $\sqrt{\omega_0^2+\omega^2}\approx \omega_0$. If we take the `oscillatory' phase of the inflaton, and use its mass to set a frequency scale, then that is related to the Planck mass and $\omega_0\sim 10^{-6} M_{\rm pl}$. 

{\it Note that this is one of the results of the paper, in the next two sections we discuss the massive scalar and the chiral fermion interactions, and the latter is more important as the results have implications for cosmic neutrino back ground and for neutrino experiments on earth.}
 
\section{\label{sec: KG massive} Klein-Gordon for a massive particle \\} 

Now, let's consider the Klein-Gordon equation for a massive particle

\begin{equation}
-\frac{1}{\sqrt{\text{-}g}}\partial_\mu \left( \sqrt{\text{-}g} \partial^\mu \phi \right)+m^2 \phi =-  \partial_{\nu} \left( g^{\mu\nu} \partial_{\mu} \phi \right) +m^2 \phi = 0 ,
\end{equation}

along with the metric of linear gravity, so $g_{\mu\nu} = \eta_{\mu\nu} + h_{\mu\nu}$ and $g \approx -1$. Expanding all the terms in the equation and using the same calculations as for the massless particle we get

\begin{equation}
0 = \DAlambert \phi -m^2\phi - \left[ h_+ ( \partial_{xx}  - \partial_{yy} ) + 2h_x \partial_y \partial_x  \right] \phi
\end{equation}

As in the previous section we can take  $h_+ = A_+ \cos(\omega(z-t))$ and $h_{\times}=A_\times \cos(\omega(z-t)+ \delta)$ and taking the first order approximation (with respect to the gravitational wave strain), we can take $\phi \approx \phi_0 + A_+ \phi_+ + A_\times \phi_\times $, where $\phi_+ = \frac{\partial }{\partial A_+}\phi$ and $\phi_\times = \frac{\partial }{\partial A_\times}\phi$. Substituting this approximation into the Klein-Gordon equation would give 
\begin{eqnarray}
\DAlambert \phi_0 -m^2(\phi_0 + A_+ \phi_+ + A_\times \phi_\times) - \left[ A_+ \cos(\omega(z-t)) (\partial_{xx} \phi_0 - \partial_{yy} \phi_0)+ 2 A_\times\cos(\omega(z-t) + \delta) \partial_x \partial_y \phi_0 \right] &&  \\ + A_+ \DAlambert \phi_+ + A_\times\DAlambert \phi_\times&=&0 \nonumber
\end{eqnarray}

Since $A_+$ and $A_\times$ are arbitrary constants we get the following set of equations
\begin{align}
\DAlambert \phi_0  - m^2 \phi_0 &= 0, \ \ \ \mbox{(unperturbed Klein-Gordon equation) } \\
\DAlambert \phi_+ - m^2 \phi_+ &= \cos(\omega(z-t)) (\partial_{xx} - \partial_{yy})\phi_0 =f(\textbf{x})\label{eqn:f}\\
 \DAlambert \phi_\times  - m^2 \phi_\times &= 2  \cos(\omega(z-t)+\delta) (\partial_{x} \partial_{y})\phi_0
\end{align} 

The simplest solution for the unperturbed Klein-Gordon equation is just a plane wave with a modified dispersion relation in Minkowski's metric: $ k_{\mu}k^\mu=-\omega^2 + \lVert \textbf{k}\rVert^2 =- m^2$. Let's take for simplicity $\phi_0$ as a plane wave with one frequency, $\omega_0$. Then the solution for the unperturbed Klein-Gordon  equation would be
\begin{equation}
\phi_0 = A_0 \mathrm{e}^{i(\textbf{k}_0 \cdot \textbf{x} - {\omega}_0 t)}
\end{equation}

In order to solve the inhomogeneous Klein-Gordon equation, we have to convert it into a momentum space through a Fourier transform. Then, if we have the equation $ \DAlambert \phi - m^2 \phi = f$ the solution is given using propagators. We first use the planar propagator to build the perturbation, and then the spherical propagator to build the perturbation. We have presented both the cases here as the results are different.  

\subsection{Planar Propagator}
\begin{equation}
G(\textbf{x}) = \frac{1}{\sqrt{(2\pi)^2}} \int_{\mathbb{R}^4} \frac{\mathrm{e}^{ik_\mu x^\mu} d^4 k}{k_\mu k^\mu + m^2}
\end{equation}
and the solution for $\phi$ would be
\be
\phi(x)= \int f (x')G (x,x') d^4 x'  
\ee
The above equation can be obtained in Fourier space as (the details can be worked out by implementing the Fourier transformation techniques as in \cite{itzykson})

\begin{equation}
\phi(x) = - \lim_{\epsilon \rightarrow 0} \int_{\mathbb{R}^4} \frac{d^4 k}{(2\pi)^4} \mathrm{e}^{i(k_\mu x^\mu)} \frac{\tilde{f}(\textbf{k})}{-k_\mu k^\mu - m^2 + i \epsilon}
\label{eqn:greenp}
\end{equation}
Where $\tilde{f}$ is the Minkowski-Fourier transform of $f$. which is given by
\begin{equation}
\tilde{f}(\textbf{k}) = \int_{\mathbb{R}^4} d^4 x \  \mathrm{e}^{-i(k_\mu x^\mu)} f(\textbf{x})
\label{eqn:fourier}
\end{equation}

Now let's calculate the integral for $f(\textbf{x}) =  \cos(\omega(z-t)) (\partial_{xx} - \partial_{yy})\phi_0 = A_0(k_{0y}^2 - k_{0x}^2)\cos(\omega(z-t)) \mathrm{e}^{ik_{0\mu}x^\mu}$ as obtained in Equation (\ref{eqn:f}). Then we find the Fourier transform of the function $f(x)$ to $\tilde f(k)$  as defined in (\ref{eqn:fourier})
\begin{align}
\tilde{f}(\textbf{k}) &= A_0 (k_{0y}^2 - k_{0x}^2) \int_{\mathbb{R}^4} d^4 x \  \mathrm{e}^{-i(k_\mu x^\mu)} \cos(\omega(z-t)) \mathrm{e}^{i k_{0\mu}x^\mu} \\
&=  A_0  (k_{0y}^2 - k_{0x}^2) \int_{\mathbb{R}^4} d^4 x \ \mathrm{e}^{-i(k_\mu x^\mu)} \mathrm{e}^{ik_{0\mu}x^\mu} \left( \frac{\mathrm{e}^{i\omega(z-t)} +\mathrm{e}^{-i\omega(z-t)}}{2}\right) \\
&= A_0 (k_{0y}^2 - k_{0x}^2) \int_{\mathbb{R}^4} d^4 x \ \mathrm{e}^{-i(\textbf{k}\cdot \textbf{x} - \omega t)} \mathrm{e}^{i(\textbf{k}_0\cdot \textbf{x} - \omega_0 t)} 
\left( \frac{\mathrm{e}^{i\omega(z-t)} +\mathrm{e}^{-i\omega(z-t)}}{2}\right) \\
&=\frac{A_0}{2} (k_{0y}^2 - k_{0x}^2) (2\pi)^4 \left( \delta(\omega - (\omega_0 + \omega))\delta(k_{0x} - k_x)\delta(k_{0y} - k_y) \delta(k_{0z} + \omega- k_z) +   \right.  \label{eqn:fou}\\
 &\left. \ \ \ \ \ \ \ \ \ \  \ \ \ \ \ \ \ \ \ \ \ \ \ \ \ \ \ \ \  \delta(\omega - (\omega_0 - \omega))\delta(k_{0x} - k_x)\delta(k_{0y} - k_y)\delta(k_{0z} - \omega - k_z)\right) \nonumber
\end{align}
Thus the solution for the inhomogeneous Klein-Gordon equation would be obtained by substituting equation (\ref{eqn:fou}) in equation (\ref{eqn:greenp})
\begin{align}
 &= -\frac{A_0}{2} (k_{0y}^2 - k_{0x}^2)  \lim_{\epsilon \rightarrow 0} \int_{\mathbb{R}^4} d^4 k \  \mathrm{e}^{i(k_\mu x^\mu)}  \cdot\\
&\cdot \frac{\left( \delta(\omega \mbox{-} (\omega_0 + \omega))\delta(k_{0x} \mbox{-} k_x)\delta(k_{0y} \mbox{-} k_y) \delta(k_{0z} + \omega \mbox{-} k_z) +  \delta(\omega \mbox{-} (\omega_0 \mbox{-} \omega))\delta(k_{0x} \mbox{-} k_x)\delta(k_{0y} \mbox{-} k_y)\delta(k_{0z} \mbox{-} \omega \mbox{-} k_z)\right)}{-k_\mu k^\mu \mbox{-} m^2 + i \epsilon} 
\nonumber \\
&=  \frac{A_0}{2} (k_{0x}^2 - k_{0y}^2)\left( \frac{e^{i((\textbf {k}_0 + \textbf{k}) \cdot \textbf {x}
- (\omega_0 + \omega) t)}}{(\omega_0 + \omega)^2 - \lVert\textbf{k}_0 + \textbf{k} \rVert^2 - m^2} + \frac{\mathrm{e}^{i((\textbf{k}_0 - \textbf{k}) \cdot \textbf{x} - (\omega_0 - \omega)t)}}{(\omega_0 - \omega_g)^2 - \lVert \textbf{k}_0 - \textbf{k} \rVert^2 - m^2} \right) \\
&=\frac{A_0}{2} (k_{0x}^2 - k_{0y}^2) \left( \frac{\mathrm{e}^{i((\textbf{k}_0 + \textbf{k}) \cdot \textbf{x} - (\omega_0 + \omega)t)}}{2 \omega (\omega_0- k_0 \cos \beta)} - \frac{\mathrm{e}^{i((\textbf{k}_0 - \textbf{k}) \cdot \textbf{x} - (\omega_0 - \omega)t)}}{2 \omega (\omega_0- k_0 \cos \beta)} \right) \label{eqn:massive}
\end{align}

Where $\cos \beta = \frac{\textbf{k} \cdot \textbf{k}_0}{ \omega \lVert \textbf{k}_0\rVert}$ and $k_0 = \lVert \textbf{k}_0 \rVert$. Note that in the above equations, the delta function integrals have been implemented for the $\tilde{f}(\textbf{k})$ obtained in equation (\ref{eqn:fou}). The scalar field $\phi(\textbf{x})$ is thus solved for as defined in equation ({\ref{eqn:greenp}) and is given explicitly by equation (\ref{eqn:massive}).
Here the solution to the massless Klein Gordon equation clearly shows two resultant waves of the scalar response, one with frequency $\omega_0+\omega$, and another with frequency $\omega_0-\omega$, as expected from the interaction of two different waves.
Note that in the way that the perturbed scalar wave is calculated, only the z-sector of the wave gets a frequency modification due to the superposition of the scalar wave and the gravitational wave in the z-direction. There are no interferences or amplitude oscillations as observed in the massless example. This is due to the fact that in the Fourier transformed space, the waves superpose as two plane waves, with no added mode components of the perturbed wave. If we use the more difficult calculation of obtaining the emergent wave using the spherical Green's function in position space, we again recover the interference pattern as obtained in the massless example.
\subsection{Spherical Propagator}
The position space of Green's function in spherical coordinates is
\be
G_m( {\textbf r}, {\textbf r'}, t-t') =\frac1{4\pi} \frac{m J_1 (m \sqrt{(t-t')^2 - |\textbf{x}- \textbf{x'}|^2})}{\sqrt{(t-t')^2 - |\textbf{x}-\textbf{x'}|^2}}
\label{eqn:green}
\ee
The above propagator is taken from Wikipedia \cite{wiki} and verified explicitly. 
Using this again gives interference patterns which could be used to detect gravitational waves on Earth. 
The perturbation would be obtained as
\be \phi_{+}\sim (k_{0x}^2 - k_{0y}^2) A_0 \int  \frac12 \left(e^{i \omega(z'-t')} + e^{-i \omega(z'-t')}\right) e^{-i \omega_0 t' + i \textbf k_0 \cdot \textbf x'} G_m( {\textbf r}, {\textbf r'}, t-t') \ d^4 x' 
\label{eqn:mass}
\ee
One the $t'$ integral is obtained, its form acquires an expression very similar to the massless case and one obtains (details given in Appendix 2).
\bea
\phi_+&\sim& \pi A_0 (k_{0x}^2-k_{0y}^2) \frac{e^{-i \tilde \omega \ t}}{||\textbf{k}_0+ \textbf{k}||} \frac{e^{i (\textbf{k}+ \textbf{k}_0)\cdot \textbf{x}}}{2\omega(\omega_0 - k_0\cos\beta)} \left[ ||\textbf {k}_0 +\textbf{k}|| \cos(||\textbf{k}_0+\textbf{k}|| t)\cos(\sqrt{\tilde\omega^2-m^2} \ t) \right. \label{eqn:spm} \\ && \left.+ \sqrt{\tilde{\omega}^2 - m^2} \sin(||\textbf{k}_0 +\textbf{k}|| \ t) \sin(\sqrt{\tilde\omega^2-m^2} \ t) - ||\textbf k_0+ \textbf{ k}||\right]  \nonumber
\eea
This is very close to the massless perturbation solution with $\tilde\omega=\omega+\omega_0$ given in equation (\ref{eqn:masslesssc}), as opposed to that for the massive scalar obtained in equation (\ref{eqn:massive}). 
The solution obtained in equation (\ref{eqn:spm}) has the same mode structure as for the massless scalar obtained in equation (\ref{eqn:masslesssc}). If we plot it, it shall show the same Oscillatory behaviour as in Figure 1.  Note that this difference in behaviour of Equation(\ref{eqn:spm}) with Equation (\ref{eqn:massive}) is in the wavefronts of the perturbations. In (\ref{eqn:massive}) we build a planar wave, and the results is that we have waves which propagate with frequencies which are sum or difference of the initial waves. This phenomena we have seen before. In (\ref{eqn:spm}) there is a change as we try to build a spherical wavefront using the spherical propagator. The result is a wave which does not have the 1/r fall off usually expected. However, we are left with a perturbation which we can interpret as a wave with a new dispersion relation $\omega= ||\bf{k} + \bf{k_0}|| \pm( \sqrt{\tilde {\omega}^2 - m^2} - \tilde{\omega})$ One can see that as $m\rightarrow 0$ the dispersion goes over to the massless case.  However, one also sees presence of frequencies of $\omega=||\bf{k} + \bf{k_0}|| \pm( \sqrt{\tilde {\omega}^2 - m^2} + \tilde{\omega})$, which is not there in the massless mode. This is plausible, as we are trying to interpret `frequencies' as in planar propagation by taking the exponentials of the sinusoidal functions. If we take $m\rightarrow 0$ in the explicit solution of (\ref{eqn:spm}) in the sinusoidal functions, then it gives us the massless solution (\ref{eqn:masslesssc}).   We have verified the calculations, and the physical reason that we have this result is the inhomogeneous differential equation. The actual plot of the field shows that the amplitude oscillates in time and space as in Figure (\ref{fig:Scalar}). Eventually, the energetic flow is perfectly plausible as the scalar field interacts with the gravitational field and exchanges energy with the passing wave.

\section{\label{sec: Dirac + } Field perturbations for massless spin $1/2$ particles \\} 
In this section we investigate the effect of the gravitational wave on a massless fermion. This is motivated from the  fact that most particles are fermions, and in particular neutrinos are almost massless fermions. We investigate the effect of a gravitational wave on a `neutrino', assumed massless in the first approximation.
 
Lets consider again a gravitational wave in the TT gauge, but with only one of its two polarizations, $+$ or $\times$. The Dirac equation is written as 
\begin{equation}
i\gamma ^{a}e_{a}^{\mu }D_{\mu }\psi -m\psi =0.
\end{equation}
Where $\gamma^a$ are a matrix representation of the Clifford algebra and these matrices satisfy the equation $\{ \gamma^a,\gamma^b \} = 2\eta^{ab}$, $e_{a}^{\mu }$ (a,b are indices in the tangent space) is the so called tetrad and satisfies the equation
\begin{equation}
g_{\mu\nu} = e^{i}_{\mu} e^{j}_{\nu} \eta_{ij},
\end{equation}
and $D_\mu$ is the covariant derivative taking into account the spin of the particles it is defined as: 
\begin{equation}
D_{\mu }=\partial _{\mu }-{\frac  {i}{4}}\omega _{{\mu }}^{{ab}}\sigma _{{ab}},
\end{equation}
where $\omega _{{\mu }}^{{ab}}$ is the spin connection symbol and $\sigma _{{ab}}={\frac  {i}{2}}\left[\gamma _{{a}},\gamma _{{b}}\right]$ is the commutator of the Dirac matrices. One way to define the spin connection symbols is through the Christoffel symbols and the derivative of the metric as follows
\begin{equation}
 {\displaystyle \omega _{\mu }^{\ ab}=e_{\nu }^{\ a} \left(\Gamma _{\ \sigma \mu }^{\nu }e^{\sigma b}+\partial _{\mu }e^{\nu b} \right)=e_{\nu }^{\ a}\Gamma _{\ \sigma \mu }^{\nu }e^{\sigma b}-e^{\nu a}\partial _{\mu }e_{\nu }^{\ b}} \label{eqn:spinc}
\end{equation}

Now let's consider the cases of $+$ and $\times $ polarization separately. Further, let's take the Weyl (Chiral) representation of the Gamma matrices, these matrices will be
\begin{equation}
{\displaystyle \gamma ^{0}={\begin{pmatrix}0&-I_{2}\\-I_{2}&0\end{pmatrix}},\quad \gamma ^{k}={\begin{pmatrix}0&\sigma ^{k}\\-\sigma ^{k}&0\end{pmatrix}},\quad \gamma ^{5}={\begin{pmatrix}I_{2}&0\\0&-I_{2}\end{pmatrix}},}
\end{equation}
Where $\sigma^k$ are the standard Pauli matrices. The details of the Weyl equation can be found in \cite{itzykson} and for curved space-time \cite{utiyama}.

In addition, to simplify the equations we are going to consider massless particles. Plugging every component into Dirac's equation in curved space for massless particles, $ i\gamma ^{a}e_{a}^{\mu }D_{\mu }\Psi = 0$,  and keeping everything up to first order with respect of the gravitational wave strain ($h$), we obtained the following systems of equations. \\

For $+$ polarization, let's say $h_{+} = \Re (a_{+} \mathrm{e}^{i\omega(z-t)}) = A_{+} \cos(\omega(z-t)) $ which is same as that in Page 5 after Equation(8) , then the tetrad can be taken as :
\begin{equation}
e^a_{\mu}\equiv\left(\begin{array}{cccc}-1&0&0&0\\0&\sqrt{1+h_{+}}&0&0\\0&0&\sqrt{1-h_{+}}&0\\0&0&0&1\end{array}\right)\label{eqn:+tetrads}
\end{equation}
The spin connections can be calculated using the tetrads of equation (\ref{eqn:+tetrads}) and their definition given in equation (\ref{eqn:spinc}), and they are found as
\begin{eqnarray}
\omega^{01}_{x} & =&  -\frac{1}{2} \left( \frac{\partial_t h_{+}}{\sqrt{1+h_{+}}} \right)    \    \     \     \    \    \   \    \   \omega^{13}_x = \frac12 \left(\frac{\partial_z h_{+}}{\sqrt{1+h_{+}}} \  \right)   \\
 \omega^{02}_{y} & = & -\frac12 \left( \frac{\partial_t h_{+}}{\sqrt{1-h_{+}}} \right)   \   \   \   \   \   \    \     \  \omega^{23}_{y} = \frac12 \left( \frac{\partial_z h_+}{\sqrt{1-h_+}} \right) 
\end{eqnarray} 

The Weyl equations are then written and separated as:
\begin{eqnarray}
\bar{\sigma}^{\mu} \partial_{\mu} \psi_L & = & \sigma^2 \frac{h_+}{2} \partial_y \psi_L - \sigma^1 \frac{h_+}{2} \partial_x \psi_L  \label{eqn:ferm1}\\
\sigma^{\mu}\partial_{\mu} \psi_R &=& \sigma^1 \frac{h_+}{2} \partial_x \psi_R - \sigma^2\frac{h_+}{2} \partial_y \psi_R \label{eqn:ferm2}
\end{eqnarray}
where $\bar\sigma^{\mu}= (1,-\pmb{\sigma})$ and $\sigma^{\mu}=(1,\pmb{\sigma})$, and $\psi_L,\psi_R$ are the two component chiral spinors such that $\psi\equiv(\psi_L,\psi_R)$. As evident from the above equations, the Fermion wave must have at least a non-zero propagation in the $x$ or $y$ direction to produce a perturbation as observed in the scalar case too.

Now, we take a perturbation $\psi=\psi_0+ A_{+} \tilde{\psi}$, and $\psi_0$  must satisfy the Weyl equation in a flat space and $A_+$ is the amplitude of the $+$ polarization of the gravitational wave. These solutions are given by the right and left handed spinors (positive and negative helicity ), $\psi_L$ and $\psi_R$ respectively. If we denote $k^{\mu}_0 = \{\omega_0, \textbf{k}_0 \}$ and $x^\mu = \{ t, \textbf{x} \}$, then the plane wave solutions for the Weyl equation in flat space are given by 

\begin{align}
\psi_{0L}(t,\textbf{x}) &= \frac{1}{\sqrt{2\omega_0(2\pi)^3}} \mathrm{e}^{ik^\mu_0 x_\mu} u^{(+)}(\textbf{k}_0) \\
\psi_{0R}(t,\textbf{x}) &= \frac{1}{\sqrt{2\omega_0(2\pi)^3}} \mathrm{e}^{ik^\mu_0 x_\mu} u^{(-)}(\textbf{k}_0),
\end{align}

where the spinors $u^{(+)}(\textbf{k}_0)$ and $u^{(-)}(\textbf{k}_0)$ are determined by the eigenvalue equations
\begin{align}
\hat{\sigma} \cdot \textbf{k}_0 u^{(+)}(\textbf{k}_0) &= -\omega_0 u^{(+)}(\textbf{k}_0) \\
\hat{\sigma} \cdot \textbf{k}_0 u^{(-)}(\textbf{k}_0) &= \omega_0 u^{(-)}(\textbf{k}_0)
\end{align}

For the next example, let's consider a particle moving in the $x$ direction. In this way $k^{\mu}_0 = \{\omega_0,\omega_0,0,0 \}$,  $u^{(+)}(\textbf{k}_0) = \frac{1}{\sqrt{2}} \left( \begin{smallmatrix} 
&1   \\ \text{-}& 1  \end{smallmatrix} \right)$ and $u^{(-)}(\textbf{k}_0) = \frac{1}{\sqrt{2}} \left( \begin{smallmatrix} 
&1   \\ &1  \end{smallmatrix} \right)$, therefore

\begin{equation}
\psi_0 = \frac{1}{\sqrt{4\omega_0(2\pi)^3}}\left( \begin{smallmatrix}
&1  \\
\text{-}&1 \\
&1  \\
&1
\end{smallmatrix} \right) \mathrm{e}^{i\omega_0 (x-t)}
\label{eqn:solution}
\end{equation}

Then, the perturbed equations are written in their separated form from equation (\ref{eqn:ferm1},\ref{eqn:ferm2}). We have added an appendix to show the explicit separation steps.

\bea
\DAlambert \tilde{\psi}_{L1} &= &\omega_0^2e^{i \omega_0 (x-t)}\cos\omega(z-t) \label{eqn:ferms1} \\
\DAlambert \tilde{\psi}_{L2 } &=  & e^{i \omega_0 (x-t)} \left(\omega_0^2 \cos (\omega  (z-t)) - i\omega\omega_0 \sin\omega(z-t) \right) \label{eqn:ferms2}\\
\DAlambert \tilde{\psi}_{R1} &= &e^{i \omega_0(x-t)}\left(\omega_0^2 \cos(\omega(z-t)) + i \omega\omega_0 \sin(\omega(z-t))\right)\\
\DAlambert \tilde{\psi}_{R2} &=  & \omega^2_0 e^{i \omega_0 (x-t)} \cos\omega  (z-t)
\eea
\begin{center}
\begin{figure}
\subfigure{\includegraphics[scale=0.3]{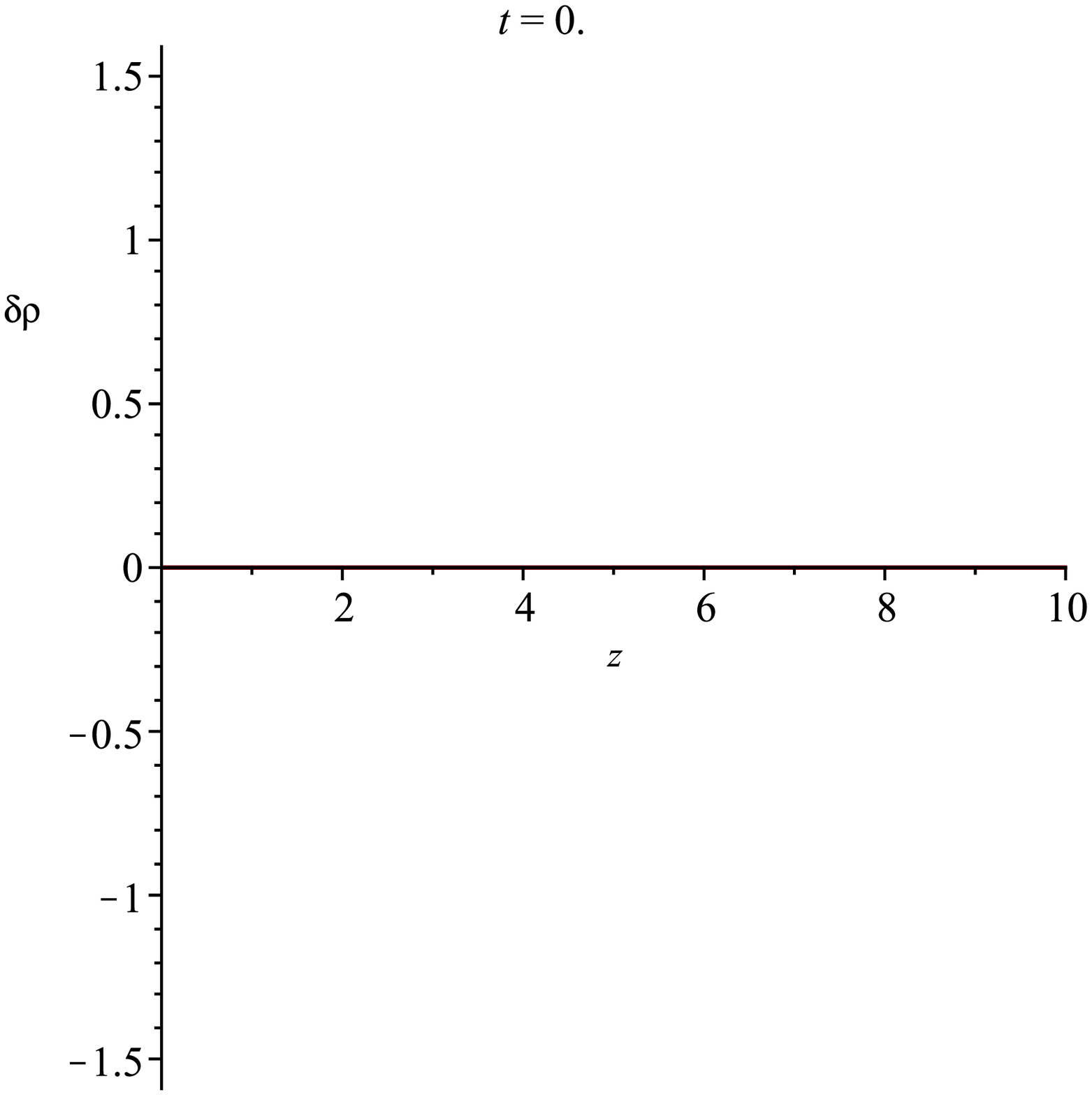}}
\subfigure{\includegraphics[scale=0.3]{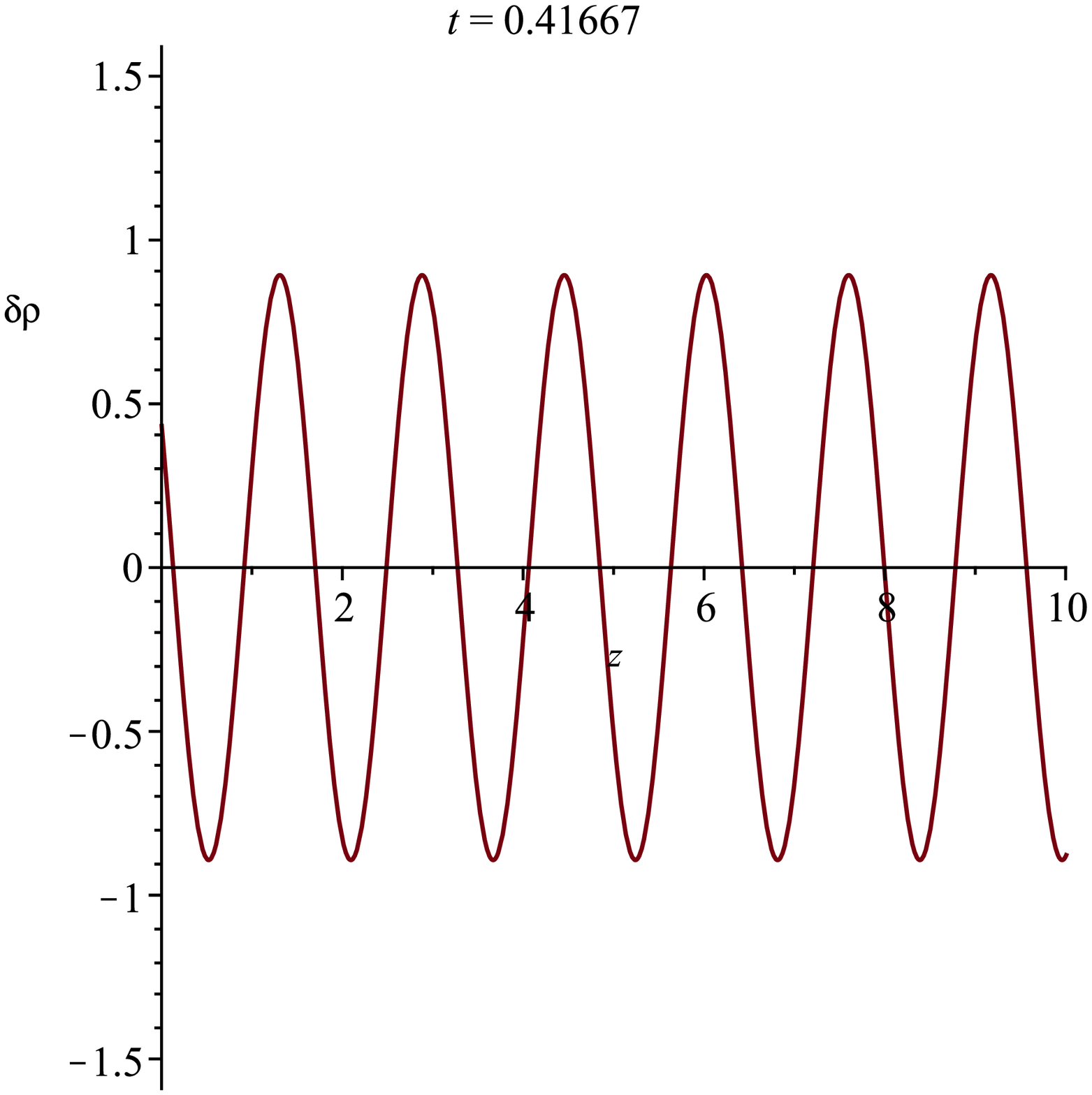}}
\subfigure{\includegraphics[scale=0.3]{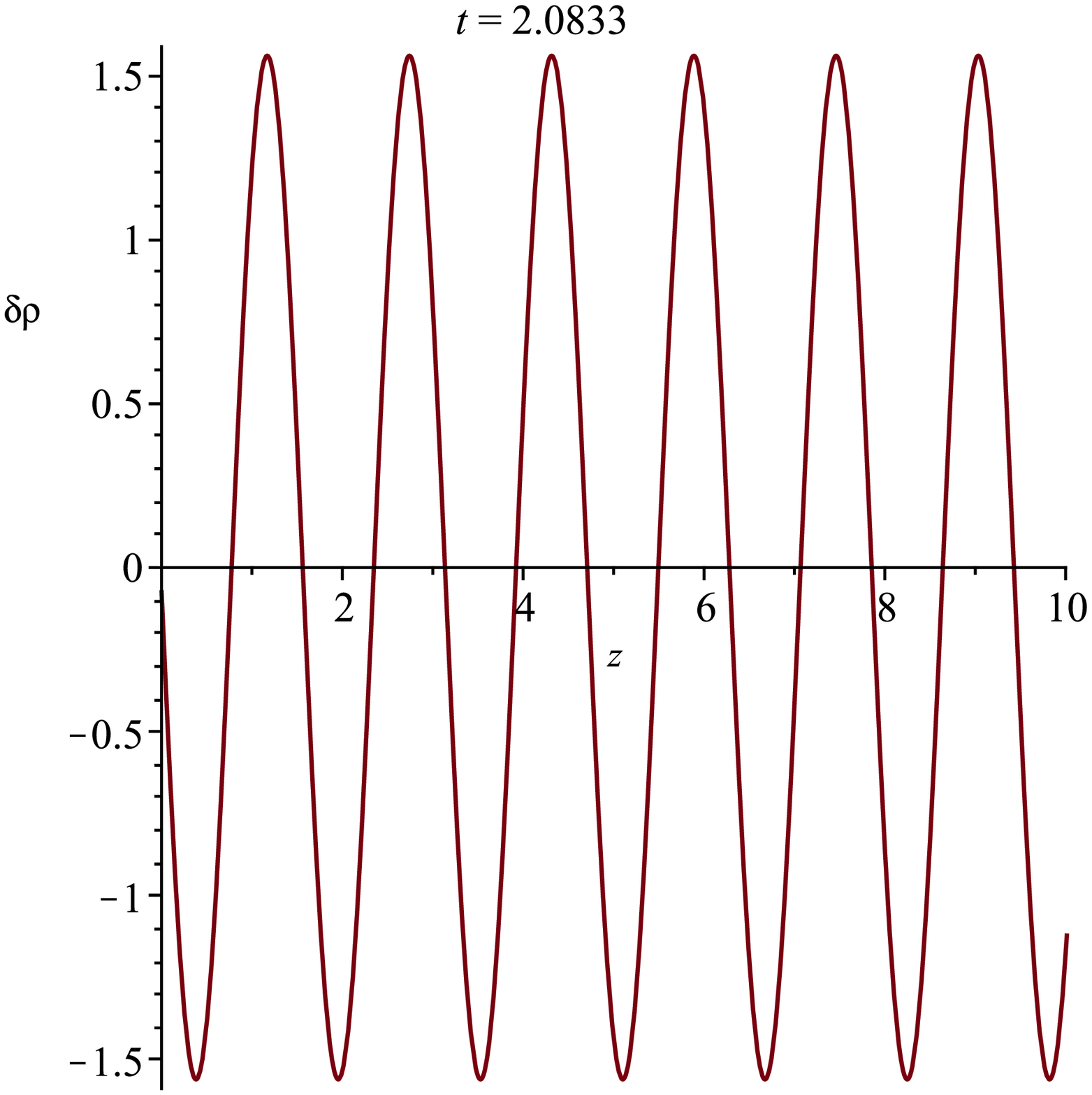}}
\subfigure{\includegraphics[scale=0.3]{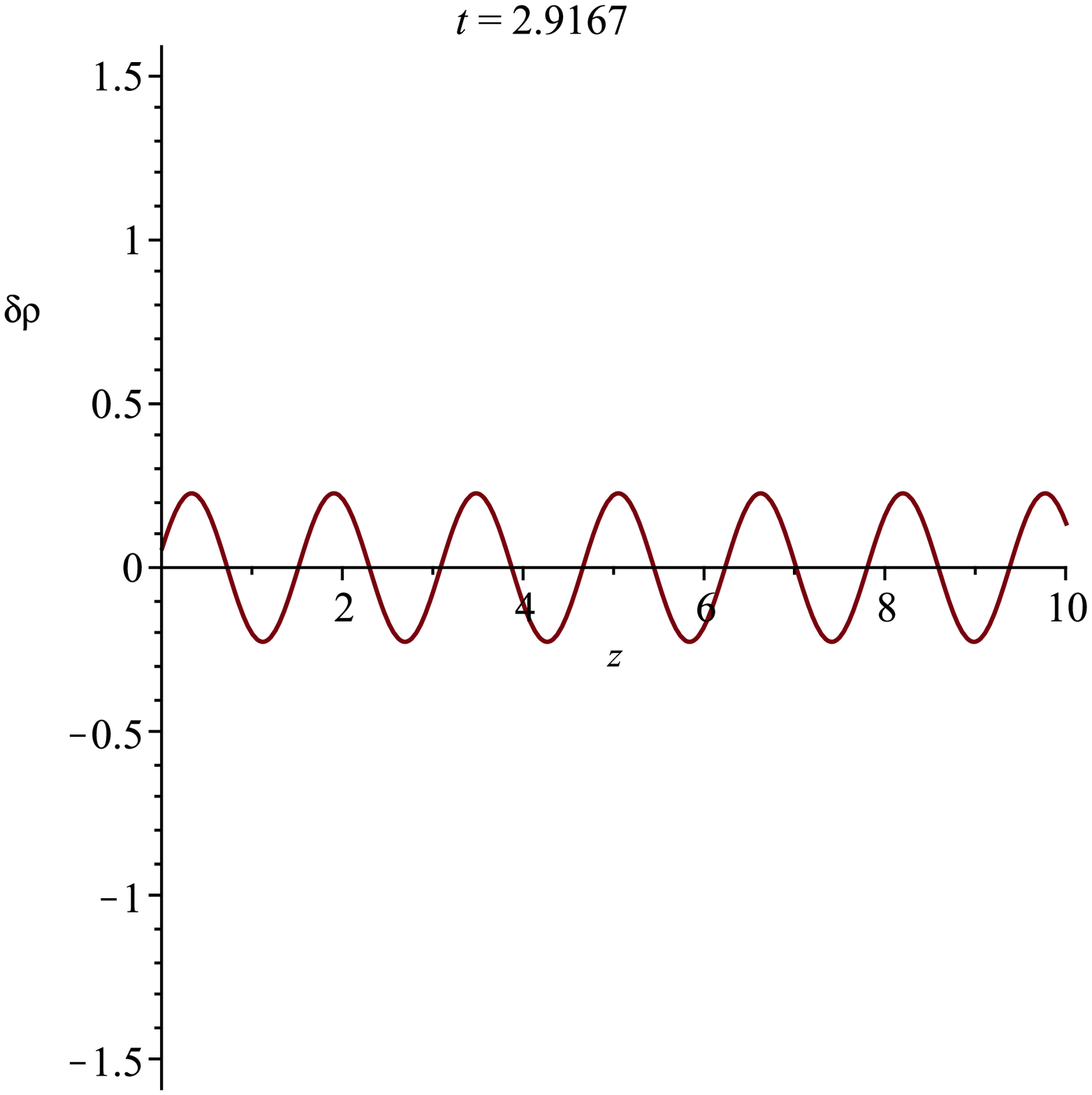}}
\subfigure{\includegraphics[scale=0.3]{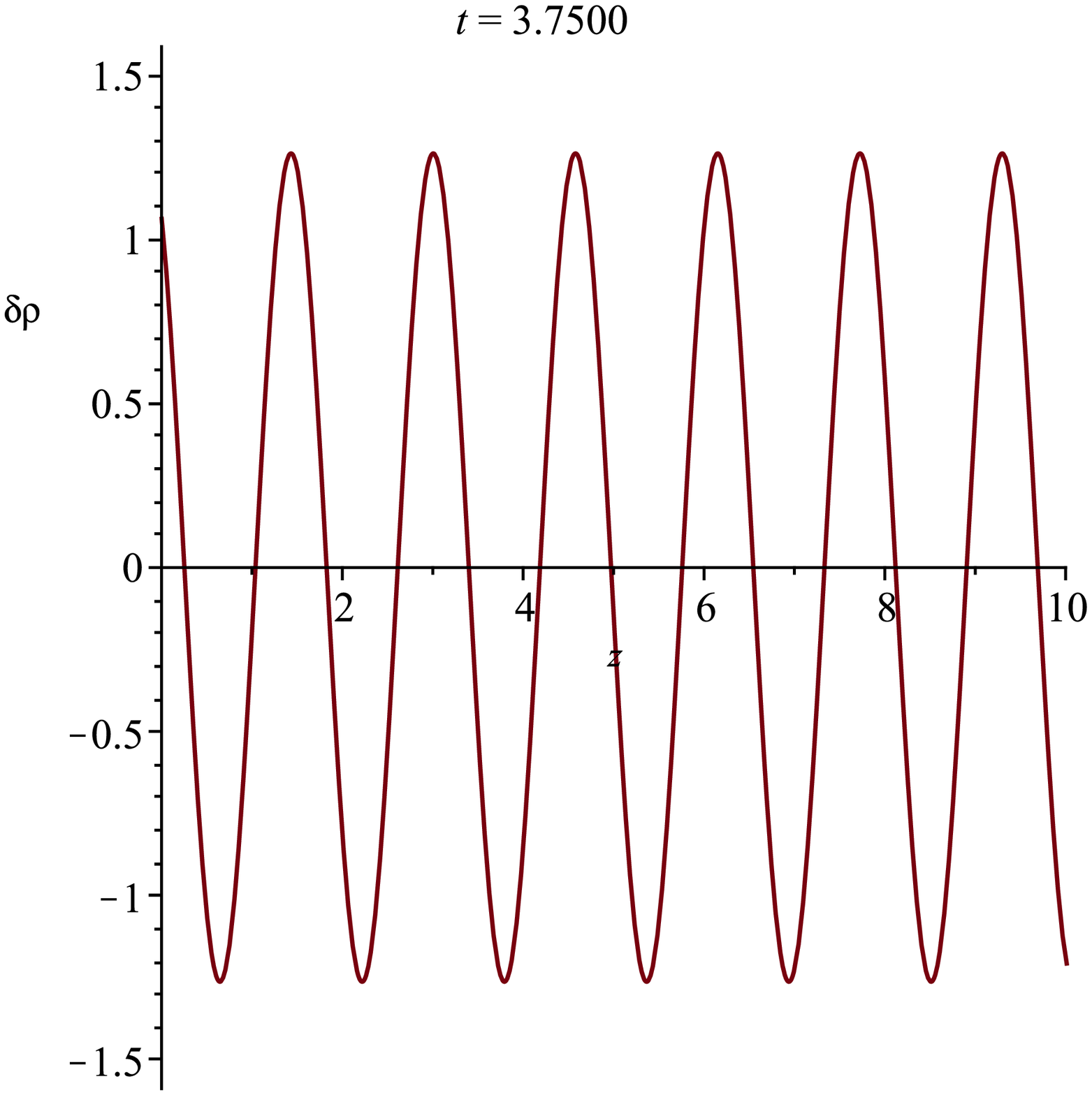}}
\subfigure{\includegraphics[scale=0.3]{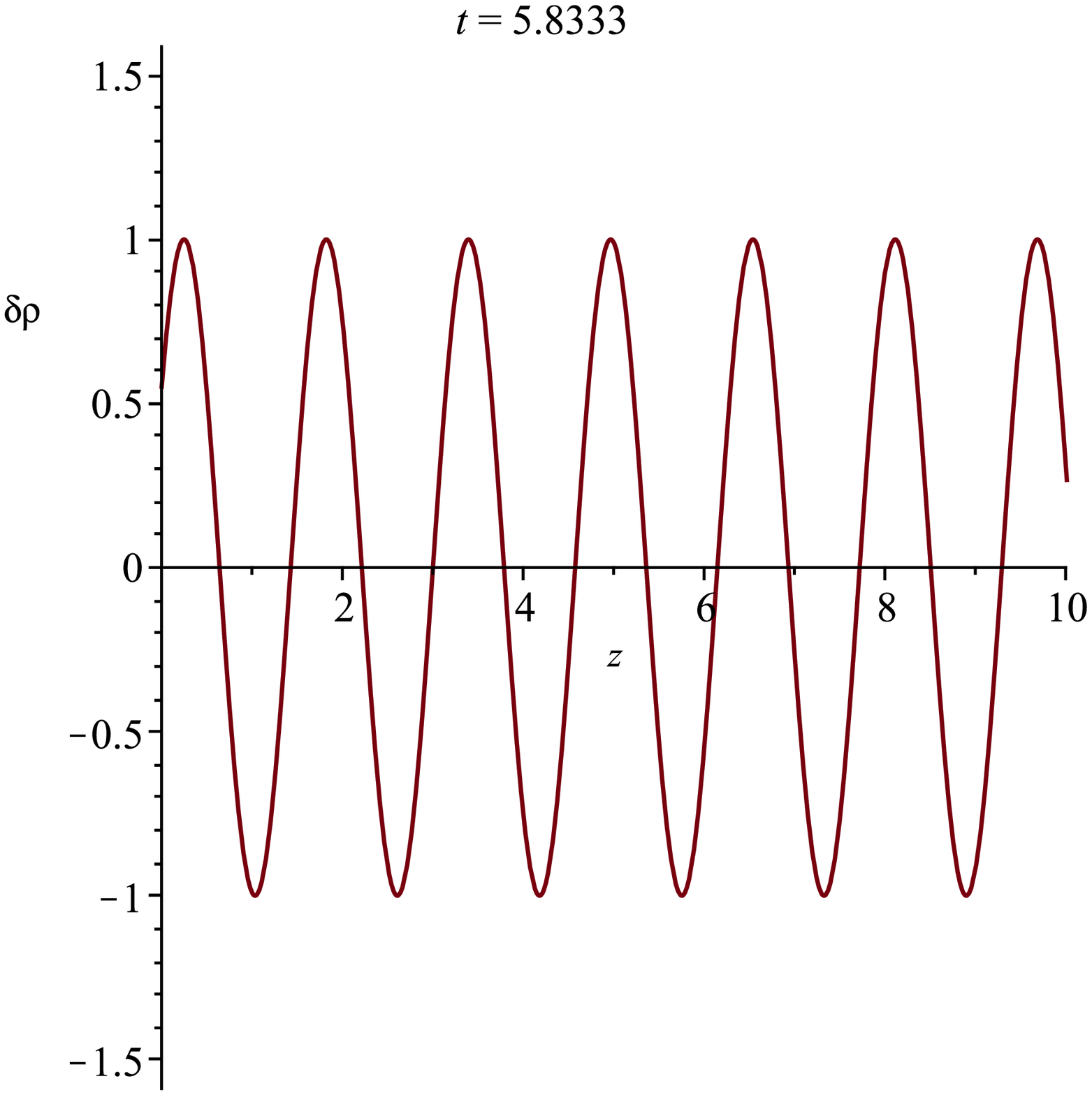}}
\subfigure{\includegraphics[scale=0.3]{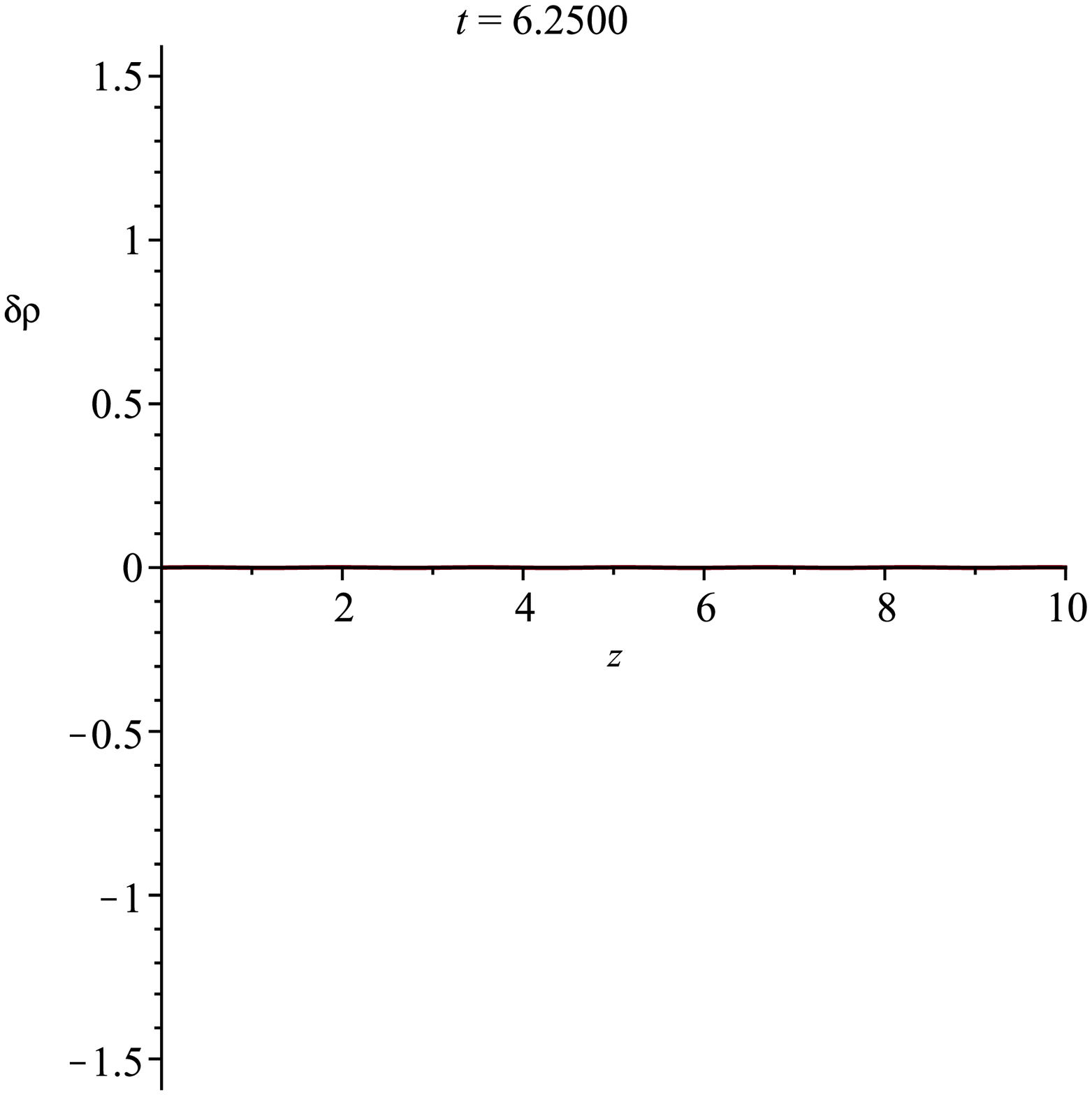}}
\subfigure{\includegraphics[scale=0.3]{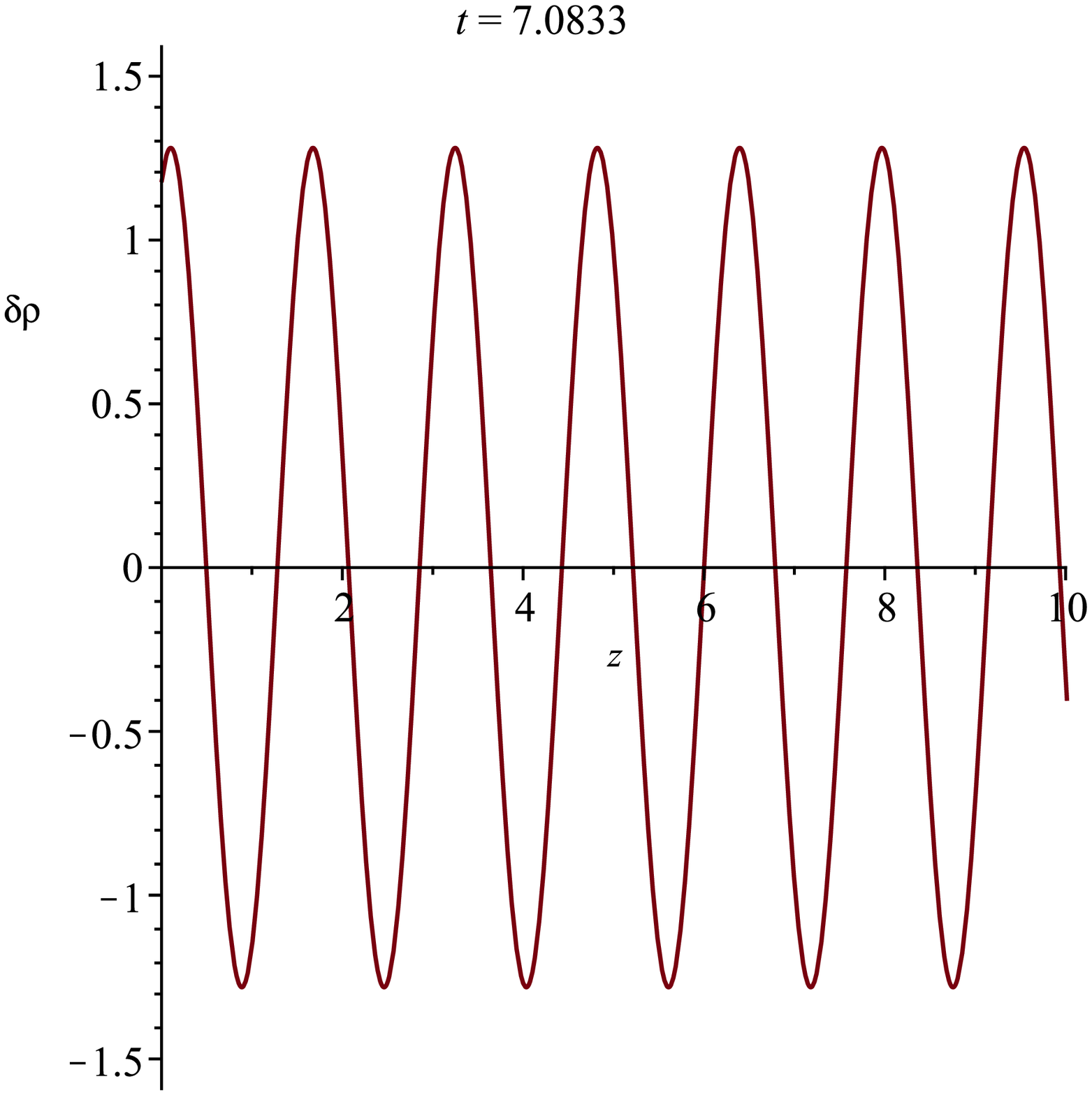}}
\subfigure{\includegraphics[scale=0.3]{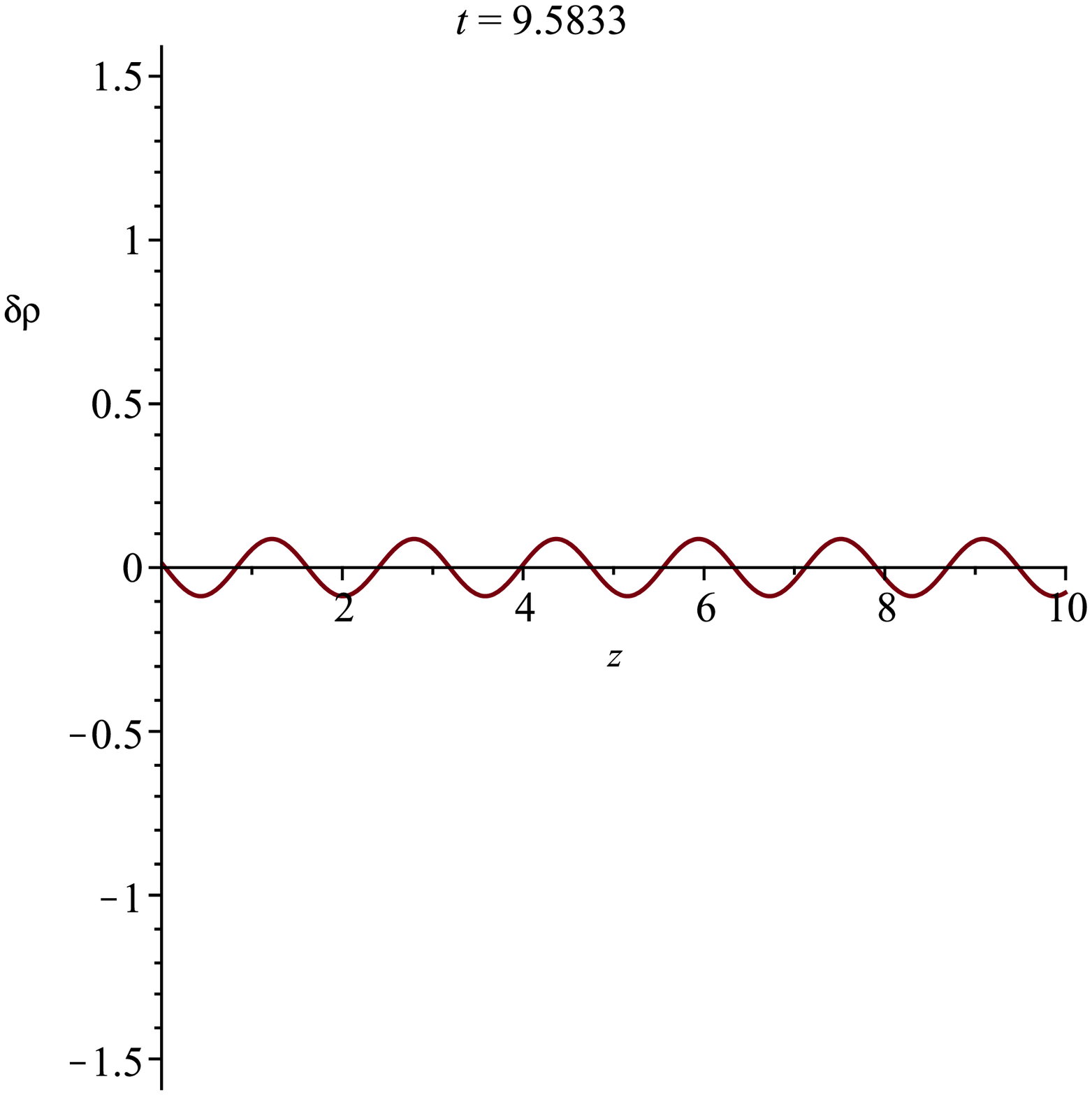}}
\caption{The correction to the Fermion density fluctuates in time and space. For the purposes of the plot $\omega=3$ and $\omega_0=4$. The plots are snapshots at 9 different time instants and are not to scale as in a real situation.}
\label{fig:Fermi}
\end{figure}
\end{center}

We write the sines and cosines in exponentials and get forms like: 

\begin{equation}
\DAlambert \tilde{\psi}_{L2 }= \frac{\omega_0}{2}e^{i \omega_0 (x-t)} \left((\omega_0-\omega) e^{i \omega(z-t)}+ (\omega_0 +\omega)e^{-i \omega(z-t)}\right)
\end{equation}
Therefore, the integral solutions of the above equations are obtained in the same way as a scalar wave perturbation. e.g.
 
\begin{eqnarray}
\tilde\psi_{L1+\omega} &= & \frac{ \pi \omega_0}{\omega}e^{i\omega_0(x-t)} e^{i\omega(z-t)}  \left[ e^{i (\omega+\omega_0) t} \cos \left(\sqrt{\omega_0^2 + \omega^2} ~t \right) + e^{i (\omega+\omega_0) t} \frac{i (\omega+\omega_0)}{\sqrt{\omega^2+\omega_0^2}}\sin\left(\sqrt{\omega^2 +\omega_0^2}~t\right) -1\right] \nn \\ 
&&\label{eqn:modesf1} \\
\tilde\psi_{L2+\omega} &=& \frac{\omega_0-\omega}{\omega_0}\psi_{L1+\omega}
\label{eqn:modesf}
\end{eqnarray}
The complete $\tilde\psi_{L1(2)}=\tilde\psi_{L1(2)+\omega}+ \tilde\psi_{L1(2)-\omega}$.

As we don't observe fermion field, we build a Fermion bilinear which represents a current: $\psi^{\dag}\gamma^0 \gamma^{\mu}\psi$, of which the 0-th component represents a `probability' density $\psi^{\dag}\psi$. This when calculated for let us say the left handed spinor gives a function which oscillates in time. 
This can be calculated as
\be
\psi^{\dag}\psi= (\psi_{0}^{\dag}+ A_{+} \tilde\psi^{\dag})(\psi_0 + A_{+} \tilde\psi) = \psi^{\dag}_0\psi_0 +  A_{+}\tilde\psi^{\dag}\psi_0 + A_{+} \psi_0^{\dag} \tilde\psi 
\ee
We compute the change in the Fermion density due to the perturbation as 
\bea
\delta\rho &= &  A_{+} \sqrt{\frac{1}{2\pi \omega_0}} \left[ \cos\omega_0 t \cos\omega z \cos(\sqrt{\omega_0^2+\omega^2} t)  + \frac{\omega}{\sqrt{\omega^2+\omega_0^2}} \cos\omega_0t \sin\omega z \cos(\sqrt{\omega^2+\omega_0^2} t)  \right. \label{eqn:den} \\ &+& \left.\frac{\omega_0}{\sqrt{\omega^2+\omega_0^2}} \sin\omega_0 t \cos\omega z \sin(\sqrt{\omega^2+\omega_0^2} t) - \cos\omega(z-t)\right] \nn
\eea

The unperturbed fermion has a constant probability density, but the 
perturbed fermion wave has an oscillatory phase. The profile of the bilinear appears  in Figure 2. The amplitude of the oscillations are proportional to $A_+$ and thus would be of the order of $10^{-21}$, compared to the initial fermion flux.

\vspace{0.5cm}

If we take the cross polarization, then, the tetrad can be taken if we take the cross polarization, then, the tetrad can be found by a matrix such that if we multiply this matrix by its transpose, we get the metric for cross polarized gravitational waves. Note that this choice of tetrads is not unique as there is internal degrees of freedom of the Lorentz rotation acting on the tangent space index \cite{utiyama}. The final Dirac equation is however invariant so any choice is equally valid \cite{utiyama}.

\begin{equation}
e_{\mu a}= \left(\begin{array}{cccc}1&0&0&0\\0&\sqrt{\frac{1-h_{\times}}{2}}&\sqrt{\frac{1-h_{\times}}{2}}&0\\0 &\sqrt{\frac{1+h_{\times}}{2}}&\sqrt{\frac{1+h_{\times}}{2}}&0\\0&0&0&1\end{array}\right) \label{eqn:Xtetrad}
\end{equation} 

The spin connections are found using the tetrads of the cross polarization given in equation (\ref{eqn:Xtetrad}) as defined in equation (\ref{eqn:spinc}) 
\be
 \omega^{12}_{t}=-\frac12 \partial_t h_{\times} 
 \ee
 
\be\omega_{x}^{01}= -\frac1{2\sqrt{2}} \partial_t h_{\times}  \   \  \   \   \  \omega_x^{02}=   -\frac1{2\sqrt{2}} \partial_t h_{\times}    \    \    \    \    \ \omega_{x}^{13} =  \frac1{2\sqrt{2}} \partial_z h_{\times}   \   \    \    \   \omega_{x}^{23} =  \frac1{2\sqrt{2}} \partial_z h_{\times} \ee
\be\omega^{01}_{y} = \frac1{2\sqrt{2}} \partial_t h_{\times} \  \   \   \   \   \   \  \omega^{02}_{y}- \frac1{2\sqrt{2}} \partial_t h_{\times}  \   \   \   \   \   \  \omega^{13}_{y} =\frac1{2\sqrt{2}} \partial_z h_{\times} \   \   \    \    \  \omega^{23}_{y} =-\frac1{2\sqrt{2}} \partial_z h_{\times}  \ee
\be\omega_{z}^{12}=-\frac1{2\sqrt{2}} \partial_z h_{\times}  \ee

The Dirac equation has two 2-component spinors written as $(\psi_L,\psi_R)$. The equations for these individual spinors are known as Weyl equations.  We write the Weyl equations with a redefinition of the coordinates $u=\frac{1}{\sqrt{2}}\left(y-x\right)$ and $v= \frac{1}{\sqrt{2}}\left(x+y\right)$;
\begin{eqnarray}
\bar{\sigma}^{\mu} \partial_{\mu} \psi_L & = & -\frac{h_{\times}}{2} \sigma^1 \partial_v \psi_L - \frac{h_{\times}}{2}\sigma^2 \partial_u \psi_L -i \frac{1}{8} \partial_t h_{\times}\sigma^3 \psi_L - \frac{3}{8} i \partial_z h_{\times}\psi_L \label{eqn:femc1} \\
\sigma^{\mu}\partial_{\mu} \psi_R &=&  \frac{h_{\times}}{2} \sigma^1\partial_v \psi_R + \frac{h_{\times}}{2} \sigma^2 \partial_u  \psi_R - i \frac{1}{8} \partial_t h_{\times} \sigma^3 \psi_R + \frac{3}{8} i \partial_z h \psi_R \label{eqn:femc2}
\end{eqnarray}

As previously, we can assume perturbative  solutions and solve the decoupled inhomogeneous equations of the D' Alembertians. Though, in this particular case, there are resonance modes as when the frequency and the direction of the Fermions and the gravitational waves coincide, there is still a non-zero inhomogeneous term. These are because of the terms independent of the derivatives of the fermion wave functions on the RHS of Equations (\ref{eqn:femc1}, \ref{eqn:femc2}).
If we assume $\psi_{L(R)}= \psi^0_{L(R)}+ A_{\times}\tilde\psi_{L(R)}$ as previously ($A_{\times}$ is the same amplitude as introduced in page 5 after equation 8) , then the perturbation equations precisely yield equations which are of the form (see Appendix for details) 
\bea
\DAlambert \tilde \psi_{L1} &= & \frac{\omega\ \omega_0}{2} \sin(\omega(z-t)) \psi^{0}_{L1} \label{eqn:femsc1}\\ 
 \DAlambert \tilde \psi_{L2} &=& - i \omega^2 \ \cos(\omega(z-t))\psi^0_{L2}  \label{eqn:femsc2}\\
\DAlambert \tilde \psi_{R1} &=&+ i \omega^2 \  \cos(\omega(z-t))  \psi^0_{R1} \label{eqn:femsc3}\\
\DAlambert \tilde \psi_{R2} &=& -\frac{\omega\ \omega_0} {2} \  \sin(\omega(z-t)) \psi^0_{R2} \label{eqn:femsc4}
\eea

We use the same solution for the unperturbed fermion as in (\ref{eqn:solution}) with $x,y\rightarrow u,v$, this is due to the different tangent frame, and solve the perturbations.  These will have corrections of the order of $\omega/\omega_0$ of the $+$ polarized gravitational waves. If we calculate the equations for the example where the directions and frequency of the two interacting waves coincide (without taking the limit $\omega\rightarrow\omega_0$ in the above), the inhomogeneous terms are proportional to $\DAlambert \tilde\psi_{L1(L2)} = -i\omega^2/2(\exp(2i\omega(z-t)) \pm 1)$ and give rise to  perturbations which grow not only linearly but quadratically, which can be interpreted as resonant modes. The explicit perturbation solutions are obtained by computing integrals (Appendix added)
\be
\tilde\psi_{L1(L2)} \sim \pi \left( i \frac{e^{2 i \omega(z+t)} - e^{2i\omega(z-t)}}{4} + \omega  t e^{2i \omega(z-t)} \mp i 2 \omega^2 t^2\right)
\label{eqn:modes}
\ee
  
  As obvious, the fermion density built from the fields will have a runaway behaviour with time as expected in resonance cases. The physical implication of this can be realized on earth looking for perturbations in a neutrino flux coming from the space, where neutrino waves are detected from a multi-messenger type event. However, the perturbation approximation might break down as the modes grow in amplitude, but definitely this behaviour is non-trivial. If we observe the bilinear it will be proportional to $A_{\times} \omega t$, so for $\omega t\sim 10^{19}-10^{20}$ the corrections will be fractions and thus can account for the neutrino background density. This is as $0 \leq A_{\times} \omega t\approx 10^{-1}  \leq \pm 1$ a fraction and the density could account for the CNB offset of $\pm 0.23$ as shown in \cite{neutspecies}. However, a careful analysis of early universe cosmology is required for this.

\section{Conclusion}
In this section we briefly summarize the results, and discuss possible implications for real physical systems (this could be a gravitational wave detector, the effect on the cosmic neutrino background, the detection of neutrino flux in a neutrino experiment from a multi messenger event etc). In this paper we studied the interaction of a gravitational wave with a scalar and a neutrino wave. In the scalar section, a new scalar perturbation mode is found,  propagating in the direction which is vector sum of the initial scalar and gravitational wave directions. The amplitude of the oscillatory scalar wave perturbation can range from $10^{-3}$ to $10^{-1}$ depending on the frequency of the gravitational waves. What this would imply for early universe cosmology is a work in progress. 
Next we examine if our results have implications for the cosmic neutrino background density. We have found for a neutrino too, a mode which propagates in a direction which is in the vector sum of the direction of propagation of the initial gravitational and neutrino wave directions. The total solution is a combination of this new mode and the usual modes which comprise of waves with sum and difference of the initial gravitational and neutrino waves.  The plot of the total perturbed neutrino density shows an oscillatory nature as shown in Figure (\ref{fig:Fermi}). To study the effect of this `scattering' of the neutrino by the gravitational wave, we have to take our calculations of the perturbation into a quantum scheme, quantizing the perturbed field and thus the interaction. By this we simply mean that the neutrino classical fields obtained will have to be written in second quantization form. This would involve introducing creation and annihilation operators. Naively though, we can calculate the change in neutrino density as $\delta \rho$ (\ref{eqn:den}) with the estimates for neutrino density in the CMB. We estimate that the flux of the neutrino will differ by a perturbation 
up to oscillatory factors. The corrections are proportional to $10^{-21}$ and thus their contribution to the effective degrees of freedom are very tiny, not in the range of the fractional corrections obtained in other papers \cite{neut1,neut2}.  In those papers non-thermal corrections to the CNB were shown to contribute to the neutrino degrees of freedom to the order of $\pm .23$. From the cross polarization though, if the gravitational waves have high frequency in the early universe, fractional corrections might appear.  At the time of neutrino decoupling the gravitational wave frequencies are quite small \cite{nuovo}, however, there can be some earlier inflationary epoch where the gravitational wave frequencies were bigger. The relic of such an interaction would propagate with a perturbation and contribute to the non-thermal fluctuations of the background neutrino effective degrees of freedom at a later time \cite{neut1} as shown in the discussions of the previous section. This is work in progress \cite{adg}. 
 
With the closed-form solutions found in this paper we could compute (numerically) second order solutions and so on. It would be interesting to compare our analysis with the one made in \cite{pprod} and look for effects such as particle production and therefore energy lost. This could explain why gravitational waves have a very small amplitude when observed on Earth and could be used to know the source's distance of gravitational waves, since the strain decays as $1/r$.

A more plausible use of our new results could be in the detection of gravitational waves on Earth in neutrino experiments. As we showed previously, a neutrino flux can be perturbed by the gravitational wave into oscillatory (note not in flavour) energy density, this is on the range of sensitivity of detectors built on Earth and thus we believe that this perturbation can be measured and compared with our calculations. As shown in (Figure \ref{fig:Fermi}), a neutrino flux gets perturbed by the gravitational wave into a scattered wave with oscillatory energy density, of the order of sensitivity of detectors built on Earth. This means that the waves are of the order $10^{-21}$ in amplitude and this has been detected in LIGO. In particular the resonant mode of the fermion perturbation for cross polarization of the gravitational wave has to be investigated further. A reference was brought to our attention about conformally coupled scalar field solutions in gravitational wave background, and implemented in cosmology \cite{kurkov}, our solutions will have implications for cosmology as well.  As neutrinos are candidate dark matter, the energy oscillations of the neutrino flux could contribute to the dark matter discussion. We hope to report on this after quantizing the neutrino flux.

\begin{acknowledgments}
{\bf Acknowledgements:} S.E.M.G. would like to thank MITACS Globalink 2017 for a summer internship fellowship. Majority of the work was completed in May-August 2017. We would like to thank Suria Itzel Morales Guzman for proofreading the manuscript. We would like to thank anonymous referee for very detailed comments and suggestions on the writing of the manuscript.
\end{acknowledgments}

\appendix
\section{Appendix 1: Integral for the massless fields}
\begin{align}
I &= \int_{\bar{B}(0,t)} \mathrm{e}^{i(\textbf{k}_0 \cdot (\textbf{x} + \textbf{r}) - \omega_0(t-r)+\delta_0)} \frac{\mathrm{e}^{i(\omega(z' - (t-r)) + \delta)}+\mathrm{e}^{-i(\omega(z' - (t-r)) + \delta)}}{2r} r^2\sin\theta d\phi d\theta dr 
\label{eqn:int}
\end{align}

Since $\textbf{x}' = \textbf{x} + \textbf{r}$ and the $z$-component of $\textbf{r}$ in spherical coordinates is $r\cos\theta$, then $z' = z + r \cos \theta$. Therefore, after taking out the independent terms in the integral, equation (\ref{eqn:define1}) is written as

\begin{align}
I &= \frac{\mathrm{e}^{i(\textbf{k}_0 \cdot\textbf{x}- \omega_0 t+\delta_0 )}}{2} \int_{\bar{B}(0,t)} \mathrm{e}^{i( \textbf{k}_0 \cdot \textbf{r} +\omega_0 r)} \left( \mathrm{e}^{i(\omega(r \cos\theta +r) + \omega(z-t)+\delta)} +\mathrm{e}^{-i(\omega(r \cos\theta +r) + \omega(z-t)+\delta)}\right) r \sin \theta d\phi d\theta dr 
\end{align}
We can always align the wave vector $\textbf{k}_0$ to be in the same direction as the $z$-axis. Then, $\textbf{k}_0 \cdot \textbf{r} = r k_0 \cos\theta$ and therefore it is $\phi$-independent. Now we can integrate the $\phi$-dependent term independently, then the $\theta$ term and finally the $r$-term as

\begin{equation}
I = \frac{\mathrm{e}^{i(\textbf{k}_0 \cdot\textbf{x}- \omega_0 t+\delta_0 )}}{2} \int_{0}^{t} \int_{0}^{\pi}  \left( \mathrm{e}^{i(\omega(z-t)+\delta)}\mathrm{e}^{i(\omega+\omega_0)r} \mathrm{e}^{i\omega r \cos\theta} +  \mathrm{e}^{-i(\omega(z-t)+\delta)}\mathrm{e}^{i(\omega_0 -\omega)r} \mathrm{e}^{-i\omega r \cos\theta}\right) \left( \int_{0}^{2\pi} \mathrm{e}^{i\textbf{k}_0 \cdot \textbf{r}} d\phi \right) r \sin\theta d\theta dr
\end{equation}

Since the only part dependant of $\phi$ is $\textbf{k}_0 \cdot \textbf{r} = k_{0x} r \sin\theta \cos \phi + k_{0y} r \sin\theta \sin \phi + k_{0z}r\cos\theta$, the integral can be easily  calculated again as a Bessel function of the first kind. Breaking again the integral into the $+$ and $-$ terms of the exponential form of the $\cos(\omega(z-t)+\delta)$, the first integral would be

\begin{align}
I_1 &= 2\pi\mathrm{e}^{i(\omega(z - t) + \delta)} \int_{0}^{t} \int_{0}^{\pi} \mathrm{e}^{i(\omega+\omega_0)r} \mathrm{e}^{i(\omega  +k_{0z} )r\cos\theta} J_0 \left( \sqrt{k_{0x}^2 + k_{0y}^2} r \sin\theta\right) r\sin\theta dr \\
&= 2\pi\mathrm{e}^{i(\omega(z - t) + \delta)} \int_{0}^{t}  \mathrm{e}^{i(\omega+\omega_0)r} \left( \int_{-1}^{
+1} \mathrm{e}^{i(\omega  +k_{0z} )r\cos\theta} J_0 \left( \sqrt{k_{0x}^2 + k_{0y}^2} r \sin\theta\right) d(\cos\theta)\right) r dr \\
&= 2\pi \mathrm{e}^{i(\omega(z - t) + \delta)} \int_{0}^{t}  \mathrm{e}^{i(\omega+\omega_0)r} \left( \int_{-1}^{+1} 2\cos((\omega  +k_{0z} )rx) J_0 \left( \sqrt{k_{0x}^2 + k_{0y}^2} r \sqrt{1-x^2}\right) dx\right) r dr \\
&\mbox{and since $k_{0x}^2+k_{0y}^2 + (k_{0z}+\omega)^2 = k_{0x}^2+k_{0y}^2 + k_{0z}^2+\omega^2 + 2k_{0z}\omega = \omega_0^2 + \omega^2 + 2k_{0z}\omega  $  }  \\
&\mbox{$\ \ \ \ \ \ \ \ \ \ \ \ \ \ \ \ \ \ \ \ \ \ \  \ \ \ \ \ \ \ \ \ \ \ \  \ \ \ \ \ \ \ = \omega_{0}^2 +\omega^2  + 2\textbf{k}\cdot \textbf{k}_0 =\lVert \textbf{k}+ \textbf{k}_0 \rVert  $, then we have} \\
&= 2\pi\mathrm{e}^{i(\omega(z - t) + \delta)} \int_{0}^{t} \mathrm{e}^{i(\omega+\omega_0)r} \left( 2\frac{\sin[\lVert \textbf{k}+ \textbf{k}_0 \rVert r]}{\lVert \textbf{k}+ \textbf{k}_0 \rVert r}\right)rdr \\
&=\frac{4\pi \mathrm{e}^{i(\omega(z - t) + \delta)}}{\lVert \textbf{k} + \textbf{k}_0\rVert} \int_{0}^t  \mathrm{e}^{i r(\omega + \omega_0)} \sin(r\lVert \textbf{k} + \textbf{k}_0\rVert) dr \\
&= \frac{4\pi \mathrm{e}^{i(\omega(z - t) + \delta)}}{\lVert \textbf{k} + \textbf{k}_0\rVert} \frac{ \left(-\lVert \textbf{k} + \textbf{k}_0\rVert+e^{i t \left(\omega +\omega _0\right)} \left(\lVert \textbf {k} + \textbf{k}_0\rVert \cos (\lVert \textbf{k} + \textbf{k}_0\rVert t)-i \left(\omega +\omega _0\right) \sin
   (\lVert \textbf{k} + \textbf{k}_0\rVert t)\right)\right)}{\left(-\lVert \textbf{k} + \textbf{k}_0\rVert+\omega +\omega _0\right) \left(\lVert \textbf{k} + \textbf{k}_0\rVert+\omega +\omega _0\right)} \\
&= \frac{4\pi \mathrm{e}^{i(\omega(z - t) + \delta)}}{\lVert \textbf{k} + \textbf{k}_0\rVert} \frac{ \left(-\lVert \textbf{k} + \textbf{k}_0\rVert+e^{i t \left(\omega +\omega _0\right)} \left(\lVert \textbf{k} + \textbf{k}_0\rVert \cos (\lVert \textbf{k} + \textbf{k}_0\rVert t)-i \left(\omega +\omega _0\right) \sin
   (\lVert \textbf{k} + \textbf{k}_0\rVert t)\right)\right)}{2\omega \omega_0 (1-\cos\beta)}
\end{align}

\appendix
\section{Appendix 2: Integral for the massive scalar}

 Isolating the $t'$ dependent 
terms in the integral (\ref{eqn:mass}) for the massive perturbation we get 
\be
 I(\mathbf{x}) = \int_{-\infty}^{(t-r)}  e^{-i (\omega+\omega_0) t'} \frac{m J_1\left(m \sqrt{(t-t')^2 -|\textbf{x}-\textbf{x'}|^2}\right)}{\sqrt{(t-t')^2 - |\textbf{x}-\textbf{x'}|^2}} dt' + (\omega \rightarrow -\omega),
 \ee
Where $I(\mathbf{x})$ is the time independent part of eq. (\ref{eqn:mass}). Writing $|\textbf{x}-\textbf{x'}| =r $ and $t-t'= \tau$, and observing that $t-t'> r$, the integral is
\be
I(\mathbf{x}) = e^{-i (\omega+\omega_0)t} \int_{r}^{\infty} \exp(-i(\omega + \omega_0)\tau) \frac{m J_1 (m \sqrt{\tau^2 - r^2})}{\sqrt{\tau^2 - r^2}} \  d\tau + (\omega\rightarrow -\omega)
\ee
In the above, we keep only the $\cos(\tilde\omega t)$ as the inclusion of the imaginary part of the exponential, the sin function gives a spurious pole at $2 m^2$ in the final answer.
\be
 = - \exp(- i\ \tilde\omega t) \  \frac{2}{r}\sin\left(\frac{r}2\left(\tilde{\omega} - \sqrt{\tilde{\omega}^2 - m^2}\right) \right)\sin\left(\frac{r}2 \left(\tilde\omega + \sqrt{\tilde{\omega}^2 -m^2}\right) \right)
 \label{eqn:time}
 \ee
 we introduce $\Lambda$ to encode the above
 \be
 =\frac2r \exp(-i \tilde\omega t) \Lambda(r, \tilde\omega)
 \ee
where $\tilde{\omega}=\omega+\omega_0.$
 The remaining exponential integrals follow the same steps as in equations (81-87).  If we label the expression in equation (\ref{eqn:time}) as $\Lambda$ then in the end the perturbation is proportional to
 \be
2 \pi~ (k_x^2-k_y^2)  e^{-i \tilde \omega t} e^{i (\textbf{k}+ \textbf{k}_0)\cdot \textbf{x}} \int_0^t \Lambda(r,\tilde\omega) \frac{\sin || \textbf{k} + \textbf{k}_0|| r}{|| \textbf{k} + \textbf{k}_0||} dr 
\ee
given as
\bea
&& 4\pi (k_{0x}^2-k_{0y}^2) \frac{e^{-i \tilde \omega \ t}}{||\textbf{k}_0+ \textbf{k}||} \frac{e^{i (\textbf{k}+ \textbf{k}_0)\cdot \textbf{x}}}{2\omega(\omega_0 - k_0\cos\beta)} \left[ ||\textbf {k}_0 +\textbf{k}|| \cos(||\textbf{k}_0+\textbf{k}||t)\cos(\sqrt{\tilde\omega^2-m^2} \  t) \right. \nn \\ && \left.+ \sqrt{\tilde{\omega}^2 - m^2} \sin(||\textbf{k}_0 +\textbf{k}|| \  t) \sin(\sqrt{\tilde\omega^2-m^2} \ t) - ||\textbf k_0+ \textbf{k}||\right]  
\eea
Similarly the correction for the transverse polarization gravitational waves can be calculated.
\appendix
\section{Appendix 3: Details of Some Derivations}
How to separate the Weyl Equations: For this we show the calculations for the Cross polarization, as that is more complicated. The $+$ polarization has a similar derivation.
The typical Weyl Equation is of the form e.g. equation (\ref{eqn:femc1}), and equation (\ref{eqn:femc2})
\bea
\bar\sigma^{\mu} \p_{\mu} \psi_L &= & f(\sigma,h_{\times},t,x,y,z)\psi_L\\
\sigma^{\mu} \p_{\mu} \psi_R &=& g(\sigma,h_{\times},t,x,y,z) \psi_R
\eea
where $ f(\sigma,h_{\times},t,x,y,z)\psi_L$ represents the RHS of Equation(\ref{eqn:femc1}) and $g(\sigma,h_{\times},t,x,y,z) \psi_R$ the RHS of Equation(\ref{eqn:femc2}).
We use the following steps:
\bea 
\left(\p_t - \vec{\sigma}\cdot \vec{\partial}\right) \psi_L & = & f(\sigma,h_{\times},t,x,y,z) \psi_L \nn \\
\left(\p_t + \vec{\sigma} \cdot \vec{\partial}\right)\left(\p_t - \vec{\sigma}\cdot \vec{\partial}\right) \psi_L&=& \left(\p_t + \vec{\sigma} \cdot \vec{\partial}\right)f(\sigma,h,t,x,y,z) \psi_L \nn \\
\left(\p_t^2 - \vec{\p}^2\right)\psi_L& = & -\frac{i}{2}  \omega^2 A_{\times} \cos(\omega(z-t))  \left(1-\sigma^3\right)\psi_L +\frac{\omega\omega_0}{8} A_{\times} \sin(\omega(z-t)) \left(\sigma^3-3\right)(-1+\sigma^1)\psi_L \nn \\ && -i\frac{\omega\omega_0}{2} A_{\times} \sin(\omega(z-t)) \sigma^2 \psi_L -\frac{A_{\times}}{2} \omega_0^2 \cos(\omega(z-t)) (\sigma^2-i \sigma^3) \nn\\
&& + \sigma^1 \frac{\omega\omega_0}{2} A_{\times} \sin(\omega(z-t)) \psi_L  \label{eqn:sep1}
\eea
In the above we have used $\{\sigma^i,\sigma^j\}=0$ and $(\sigma^{i})^2=1$ in showing that $\left(\p_t + \vec{\sigma} \cdot \vec{\partial}\right)\left(\p_t - \vec{\sigma}\cdot \vec{\partial}\right) \psi_L= (\partial_t^2 - \vec\p^2) \psi_L$ where the derivative operator becomes the diagonal Laplacian. The RHS of the above is obtained by straightforward computation.
We then assume that the wave function has a perturbation  $\psi=\psi_0+ A_{\times} \tilde{\psi}$. Using the solution for $\psi_0$ as stated in Equation (\ref{eqn:solution}), we simplify equation (\ref{eqn:sep1}) and we get the Equations (\ref{eqn:femsc1}, \ref{eqn:femsc2}) to order $A_{\times}$. 

Exactly in the same way, if we start from Equations (\ref{eqn:femc1}) and (\ref{eqn:femc2}) and let the $\psi_0$ be a wave in the z-direction, such that 
$\psi_0 \propto \exp(i\omega_0(z-t))$, we get
\be
 \left(\p_t^2 - \vec{\p}^2\right)\psi_L=\frac{-i}{4}\left[ \omega^2 A_{\times} \cos(\omega(z-t)) (\sigma^3 +1 )\psi_L + i \omega \omega_0 A_{\times} \sin(\omega(z-t)) (\sigma^3 -1) \psi_L\right]
 \label{eqn:reso}
 \ee
 

If we simplify the matrices and use the solution for $\psi_0$ as being a wave in the same direction as the gravitational wave and with the same frequency   then the RHS of (\ref{eqn:reso}) will have (upto factors) terms proportional to 
$\sin(\omega (z-t))\exp(i\omega(z-t))\approx \exp(2i \omega(z-t))- 1$ and $\cos(\omega (z-t))\exp(i\omega(z-t))\approx \exp(2i \omega(z-t))+1$ and the integral of this yields the perturbations as Equations (\ref{eqn:modes}). 
The integral is formulated exactly as in the previous calculations as in Equation(\ref{eqn:int})
\be
I= \int_{\bar {B}(0,t)} \frac{e^{2 i \omega(z+r \cos\theta -(t-r))} \pm 1}{r}  r^2 \sin\theta dr d\theta d\phi
\ee
A step by step integration yields Equation (\ref{eqn:modes}).

\end{document}